\PassOptionsToPackage{unicode}{hyperref}
\PassOptionsToPackage{hyphens}{url}
\PassOptionsToPackage{dvipsnames,svgnames,x11names}{xcolor}
\documentclass[
  12pt]{article}

\usepackage{amsmath,amssymb}
\usepackage{iftex}
\ifPDFTeX
  \usepackage[T1]{fontenc}
  \usepackage[utf8]{inputenc}
  \usepackage{textcomp} % provide euro and other symbols
\else % if luatex or xetex
  \usepackage{unicode-math}
  \defaultfontfeatures{Scale=MatchLowercase}
  \defaultfontfeatures[\rmfamily]{Ligatures=TeX,Scale=1}
\fi
\usepackage{lmodern}
\ifPDFTeX\else  
\fi
\IfFileExists{upquote.sty}{\usepackage{upquote}}{}
\IfFileExists{microtype.sty}{% use microtype if available
  \usepackage[]{microtype}
  \UseMicrotypeSet[protrusion]{basicmath} % disable protrusion for tt fonts
}{}
\makeatletter
\@ifundefined{KOMAClassName}{% if non-KOMA class
  \IfFileExists{parskip.sty}{%
    \usepackage{parskip}
  }{% else
    \setlength{\parindent}{0pt}
    \setlength{\parskip}{6pt plus 2pt minus 1pt}}
}{% if KOMA class
  \KOMAoptions{parskip=half}}
\makeatother
\usepackage{xcolor}
\makeatletter
\ifx\paragraph\undefined\else
  \let\oldparagraph\paragraph
  \renewcommand{\paragraph}{
    \@ifstar
      \xxxParagraphStar
      \xxxParagraphNoStar
  }
  \newcommand{\xxxParagraphStar}[1]{\oldparagraph*{#1}\mbox{}}
  \newcommand{\xxxParagraphNoStar}[1]{\oldparagraph{#1}\mbox{}}
\fi
\ifx\subparagraph\undefined\else
  \let\oldsubparagraph\subparagraph
  \renewcommand{\subparagraph}{
    \@ifstar
      \xxxSubParagraphStar
      \xxxSubParagraphNoStar
  }
  \newcommand{\xxxSubParagraphStar}[1]{\oldsubparagraph*{#1}\mbox{}}
  \newcommand{\xxxSubParagraphNoStar}[1]{\oldsubparagraph{#1}\mbox{}}
\fi
\makeatother

\usepackage{longtable,booktabs,array}
\usepackage{calc} % for calculating minipage widths
\usepackage{etoolbox}
\makeatletter
\patchcmd\longtable{\par}{\if@noskipsec\mbox{}\fi\par}{}{}
\makeatother
\IfFileExists{footnotehyper.sty}{\usepackage{footnotehyper}}{\usepackage{footnote}}
\makesavenoteenv{longtable}
\usepackage{graphicx}
\makeatletter
\def\maxwidth{\ifdim\Gin@nat@width>\linewidth\linewidth\else\Gin@nat@width\fi}
\def\maxheight{\ifdim\Gin@nat@height>\textheight\textheight\else\Gin@nat@height\fi}
\makeatother
\setkeys{Gin}{width=\maxwidth,height=\maxheight,keepaspectratio}
\makeatletter
\def\fps@figure{htbp}
\makeatother

\makeatletter
\@ifpackageloaded{caption}{}{\usepackage{caption}}
\AtBeginDocument{%
\ifdefined\contentsname
  \renewcommand*\contentsname{Table of contents}
\else
  \newcommand\contentsname{Table of contents}
\fi
\ifdefined\listfigurename
  \renewcommand*\listfigurename{List of Figures}
\else
  \newcommand\listfigurename{List of Figures}
\fi
\ifdefined\listtablename
  \renewcommand*\listtablename{List of Tables}
\else
  \newcommand\listtablename{List of Tables}
\fi
\ifdefined\figurename
  \renewcommand*\figurename{Figure}
\else
  \newcommand\figurename{Figure}
\fi
\ifdefined\tablename
  \renewcommand*\tablename{Table}
\else
  \newcommand\tablename{Table}
\fi
}
\@ifpackageloaded{float}{}{\usepackage{float}}
\floatstyle{ruled}
\@ifundefined{c@chapter}{\newfloat{codelisting}{h}{lop}}{\newfloat{codelisting}{h}{lop}[chapter]}
\floatname{codelisting}{Listing}

\makeatother
\makeatletter
\@ifpackageloaded{caption}{}{\usepackage{caption}}
\@ifpackageloaded{subcaption}{}{\usepackage{subcaption}}
\makeatother

\ifLuaTeX
  \usepackage{selnolig}  % disable illegal ligatures
\fi
\usepackage[]{natbib}
\usepackage{bookmark}

\IfFileExists{xurl.sty}{\usepackage{xurl}}{} % add URL line breaks if available
\hypersetup{
  pdftitle={Title},
  pdfauthor={Author 1; Author 2},
  pdfkeywords={3 to 6 keywords, that do not appear in the title},
  colorlinks=true,
  linkcolor={blue},
  filecolor={Maroon},
  citecolor={Blue},
  urlcolor={Blue},
  pdfcreator={LaTeX via pandoc}}

\newcommand{\anon}{1}

\usepackage{amsfonts, mathtools}
\usepackage{enumerate}
\usepackage[inline]{enumitem} 
\usepackage{times}
\usepackage{booktabs}
\usepackage{url}
\usepackage{color,soul}
\usepackage{listings}
\usepackage{tocloft}
\usepackage{algorithm}
\usepackage{algpseudocode}
\usepackage{subcaption} %subfigures
\usepackage{placeins} %float barriers

\newcommand{\expfunc}[1]{\exp\left\{  #1 \right\}}

\newcommand{\indep}[0]{\perp\kern-5pt\perp}

\DeclareMathOperator*{\argmin}{arg\,min}

\begin{document}

\def\spacingset#1{\renewcommand{\baselinestretch}%
{#1}\small\normalsize} \spacingset{1}

%%%%%%%%%%%%%%%%%%%%%%%%%%%%%%%%%%%%%%%%%%%%%%%%%%%%%%%%%%%%%%%%%%%%%%%%%%%%%%

\if1\anon
{
  \title{\bf Dynamic Co-Expression Network Estimation via Multivariate Mixed-Effects Models}

  \author{Samuel Ozminkowski \\
    Department of Statistics and Data Science, Northwestern University\\
    Lifang Hou \\ 
    Department of Preventative Medicine, Feinberg School of Medicine, \\ Northwestern University \\
    David R. Jacobs, Jr. \\
    Division of Epidemiology and Community Health, School of Public Health, \\ University of Minnesota \\ 
    and \\
    Hongmei Jiang \\
    Department of Statistics and Data Science, Northwestern University}
  \maketitle
} \fi

\if0\anon
{
  \bigskip
  \bigskip
  \bigskip
  \begin{center}
    {\LARGE\bf Dynamic Co-Expression Network Estimation via Multivariate Mixed-Effects Models}
\end{center}
  \medskip
} \fi

\bigskip
\begin{abstract}
High-throughput sequencing technologies have enabled the collection of large-scale longitudinal -omics data, providing new opportunities for studying co-expression networks among molecular nodes such as genes and proteins. However, the high dimensionality and temporal dependence inherent in such data require specialized statistical methods. We propose a novel approach to infer dynamic co-expression networks among features over time (DCENt), where each node (feature) is modeled with a mixed-effects model, and dependencies among nodes are captured through correlated random effects. We develop two innovative penalized algorithms which harness the state of the art of threshold covariance estimators to estimate the random-effects covariance structure. Simulation studies show improved performance over existing approaches in terms of both mean square error and mean absolute error. We further apply the methods to data from the CARDIA study to investigate how the protein co-expression networks evolve over time as well as the association between protein trajectory patterns.
\end{abstract}

\noindent%
{\it Keywords:} Mixed-models, dynamic co-expression network, thresholding estimator, covariance estimation 
\vfill

\newpage
\spacingset{1.8} % DON'T change the spacing!

\section{Introduction}

The last decades have seen an explosion in research into human -omics such as microbiomics and proteomics, driven largely by advances in high-throughput sequencing technology \citep{prohaska_use_2011}. For example, in medical practice, protein levels are often used to diagnose disease, and many studies have been conducted to find associations between protein abundance and disease \citep{greber_decreased_1999,harper_proteome_2016,erez_prediction_2017,tarca_prediction_2019,deshmukh_deep_2021,libiger_longitudinal_2021,kandpal_gut_2024}. Despite the large number of existing studies, there is still work to be done improving methods for these areas due to the complexity of the data. In -omics data, the relationships between nodes such as proteins, genes, or other features are generally believed to be sparse - most nodes are not connected to most other nodes \citep{saeys_review_2007}. For cross-sectional data, many methods have been proposed to study the co-expression networks between these nodes. For example, for the proteome, \citet{hasman_uncovering_2023} provide a thorough overview of current tools for protein co-expression networks such as WGCNA, MIDER, and SEC \citep{langfelder_wgcna_2008,villaverde_mider_2014,cui_sparse_2016a}. Another great overview of protein co-expression network estimation is given in \citet{vella_proteinprotein_2017}. Of these, WGCNA is the most popular tool. Though the G stands for gene, it can be used with any -omics data to estimate co-expression networks.

Each of these methods utilizes the assumption that only one sample is taken from each subject. When longitudinal data is used, where multiple samples are taken from each subject over a period of time, new methods are required. Mixed-effects methods allow for more complex variance structures than ordinary least squares regression models, and are thus particularly useful for modeling longitudinal relationships \citep{laird_randomeffects_1982}. These models allow handling of irregular and missing timepoints naturally. They can be used for prediction and to estimate covariance between proteins or genes. These models can be fit via (restricted) maximum likelihood methods, such as with the popular R package lme4 \citep{bates_fitting_2015a}. However, adding random effects also introduces new complexities, especially when data is high dimensional ($p \gg n$), which is typical in -omics data \citep{saeys_review_2007}.

Various forms of high-dimensional mixed effects models have been studied in the literature. Broadly speaking, these methods can be split into three categories: methods that penalize the fixed effect coefficients but assume the random effects are not high dimensional \citep{schelldorfer_estimation_2011a,ghosh_nonconcave_2018,gorstein_highdimmixedmodelsjl_2025}, methods that penalize the random effect coefficients but assume the fixed effects are not high dimensional \citep{huang_covariance_2006,bondell_joint_2010, wang_doubly_2010, peng_model_2012}, and methods that penalize both fixed and random effects coefficients \citep{krishna_shrinkagebased_2008, ibrahim_fixed_2011a,fan_variable_2012,lin_fixed_2013,li_doubly_2018a}. The most similar work to ours comes from \citet{ahn_momentbased_2012}, who present two methods for univariate random effect selection in mixed models. However, these methods differ from ours in a few key ways. First and foremost, each of these methods is focused on univariate models, whereas we utilize a system of models for a multivariate outcome, Furthermore, each of these methods only penalize the diagonal elements of the covariance matrix. Our method is able to penalize both the diagonal and off-diagonal elements of the covariance matrix, which is pivotal in the -omics data space. This gets to the crux of the matter as far as this paper is concerned. 

\citet{fieuws_joint_2004a} first introduced a joint modelling framework using mixed effects models for the case where there are only two nodes. Here, each node is described by the linear mixed model. The bivariate linear mixed model is constructed by linking the two nodes via a joint distribution of random effects. Expanding into an arbitrary number of nodes, \citet{fieuws_pairwise_2006} introduced the pairwise estimation technique for high dimensional data. Instead of maximizing the log-likelihood of the joint mixed model, log-likelihoods of this form are maximized separately. In this paper, we propose new algorithms to estimate the interactions among all nodes simultaneously.

The major difference between our work and previous mixed-effects estimation methods is that we are not focused solely on variable selection. Rather, we propose a system of mixed-effects models to infer dynamic co-expression networks. In our models, each node is modeled with a random intercept (subject-specific variability) and a random slope (temporal variability), and dependencies among nodes are captured through correlated random effects. Our method focuses on estimating the variance-covariance matrix of the random effects, which is assumed to be sparse with a large number of nodes, for the multivariate mixed-effects model. In this context, sparse refers to the presence of many zero pairwise associations, rather than any specific block-diagonal or modular structure. The goal of this paper is to estimate the parameters of this joint multivariate mixed-effects model so that the correlations between trajectory patterns and marginal correlations between nodes over time can be derived. We introduce two algorithms for mixed model estimation with high-dimensional random effects utilizing thresholding methods for covariance matrix sparsification. Through simulation studies we show the strengths and weaknesses of each algorithm, and show that both outperform pairwise estimation in terms of both MSE and $L_1$ error. Both algorithms are much faster than the pairwise approach (using lme4), especially at high dimension and/or large sample size. Finally, we apply these methods to proteomics data from the CARDIA study to investigate how the protein co-expression networks evolve over time as well as the association between protein trajectory patterns. R code for both algorithms is available on github at \url{https://github.com/samozm/DCENt}.

\section{Methods}\label{sec:mixeff}

\begin{figure}[b!]
    \centering{
    \includegraphics[width=6in,height=\textheight]{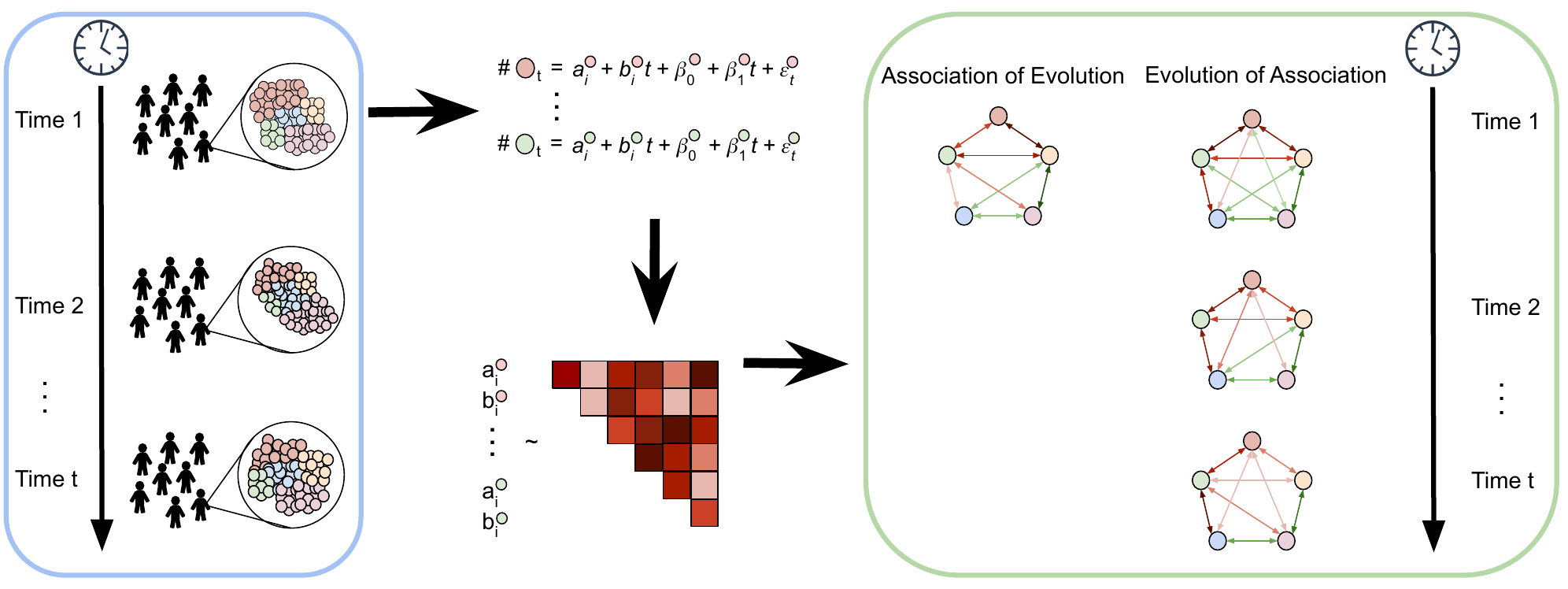}}
    \caption{The pipeline starts by obtaining measurements of multiple features (e.g., proteins or genes) from the same group of subjects repeatedly over time. A multivariate mixed-effects model is fit to the data and the covariance matrix between the random effects is estimated. These estimates characterize the (time-independent) association of evolution and the (time-dependent) evolution of association.}
    \label{fig:graphical_abstract}
\end{figure}

An overview of the proposed work is presented in Figure \ref{fig:graphical_abstract}. The data are longitudinal in nature, consisting of repeated measurements of multiple features, such as proteins, collected from the same set of subjects over time. Throughout this paper, we use the terms features and nodes interchangeably. A multivariate mixed-effects model is employed to model these data, allowing for subject-specific variability and the dependence structure among features. Based on the estimated covariance matrix of the random effects, we quantify two aspects of dependence: the association of feature trajectories, reflecting the association among their slopes, and the evolution of association, which captures how the relationships between features vary over time. Before introducing the proposed model and algorithms, we briefly review the relevant literature on covariance estimation.

\subsection{Covariance estimation}

A good overview of unconstrained covariance estimation is provided by \citet{fan_overview_2015}. One of the most important ideas in large covariance matrix estimation is thresholding, first presented by \citet{bickel_covariance_2008}. In this framework, some cells in the covariance matrix are systematically reduced or set to zero. The most basic form of thresholding is hard thresholding, where all values below the threshold in the matrix are set to zero. This could be done by setting one threshold for the entire matrix, as in \citet{bickel_covariance_2008} or it could be done by adaptive thresholding, which takes into account that each cell in the matrix may have different scales. These methods use the standard error of each cell to set different cell-specific thresholds \citep{cai_adaptive_2011}. Generalized thresholding methods smooth the thresholding function around the threshold. These methods generally follow basic rules established by \citet{antoniadis_regularization_2001}. Thresholds of this type include SCAD \citep{fan_variable_2001} and MCP \citep{zhang_nearly_2010}. Further adaptive thresholding methods have been presented by \citet{lam_highdimensional_2020}, \citet{fan_large_2013}, and \citet{fang_adaptive_2024}. The major benefit of this type of threshold is that the estimates produced from these methods tend to be continuous in the maximum likelihood estimates, which is not true when hard thresholding is used \citep{fan_overview_2015}.

\subsection{Multivariate mixed-effects model}

\citet{fieuws_joint_2004a} first introduced a joint modelling framework using mixed effects model for a longitudinal study with only two nodes. Here, each node is described by a linear mixed model. We follow their subsequent work \citep{fieuws_pairwise_2006} and extend the model to an arbitrary number of nodes. For each node $j=1,...,K$, we have the linear mixed effects model
\begin{align}\label{eq:mixeff}
    Y_{ij}(t) &= \alpha_j + a_{ij} + \gamma^\top W_i + (\beta_j + b_{ij})t + \varepsilon_{ijt}
\end{align} 
\begin{itemize}
    \item $Y_{ij}(t)$ is the measurement of the $j$th node of subject $i$ at time $t$.
    \item $\alpha_j$, $\beta_j$ are the fixed intercept and slope associated with node $j$ (respectively). Denote the vector $\boldsymbol\beta^\top = (\alpha_1,\beta_1,...,\alpha_K,\beta_K)$,
    \item $a_{ij}$, $b_{ij}$  are the random intercept and slope associated with node $j$ for subject $i$ (respectively). Denote the vector $\textbf{b}^\top = (a_{11}, b_{11}, ..., a_{1K}, b_{1K}, a_{21}, b_{21}, ... , a_{NK}, b_{NK})$.
    \item $W_i$ is an optional row-vector of additional fixed effect factors for subject $i$, with $\gamma^\top$ being the column-vector of corresponding coefficients.
\end{itemize}

The full model is constructed by linking the system of $K$  models via a joint distribution of random effects (Eq. \ref{eq:joint_dist}) plus error terms $\varepsilon_{ijt} \sim N(0, \sigma_j^2)$ independent of each other and of the random effects.
{\renewcommand*{\arraystretch}{0.5}
\begin{align}\begin{bmatrix}
a_{i1} \\
b_{i1} \\
 \vdots \\
a_{iK} \\
b_{iK}
\end{bmatrix} \sim N\left( \mathbf{0}, D \right), \hspace{8mm} &D = \begin{pmatrix}
\sigma_{a_{1}}^{2} & \sigma_{a_{1}b_{1}} & \ldots & \sigma_{a_{1}b_{K}} \\
 & \ddots & & \\
 & & \ddots & \\
 & & & \sigma_{b_{K}}^{2}
\end{pmatrix}, \hspace{3mm} \nonumber \\  \begin{bmatrix} \varepsilon_{i1t} \\ \vdots \\ \varepsilon_{iKt}\end{bmatrix} \sim N(0, E), \hspace{8mm} &E = \begin{pmatrix}
\sigma_{1} & & \textbf{0} \\ & \ddots \\ \textbf{0} & & \sigma_{K}
\end{pmatrix}.\label{eq:joint_dist}\end{align}}

The full covariance matrix for $Y$ is defined as $V \coloneqq Z D Z^\top  + E$, where $Y$ contains the protein (or other outcome) measurements vectorized in a nested order, with observations arranged so that all timepoints for node 1 for subject 1 appear first, then all timepoints for node 2 for subject 1 appear, and so on, this pattern repeating for each subject. That is, we model $Y \sim N(X\beta,V)$. We define Z as the random effect model matrix and $X$ as the fixed effect model matrix. However, for ease of explanation we ignore the additional fixed effects factors $W$. Denoting $T_i$ the number of timepoints for subject $i$ and $[\mathbf{T_i}] \coloneqq [1,...,T_i]^\top$, we have
{\renewcommand*{\arraystretch}{0.75}
\[Z = \begin{bmatrix}\textbf{1}_{T_1} & \mathbf{[T_1]} & & \textbf{0}\\ & \ddots & \ddots \\ \textbf{0} & & \textbf{1}_{T_n} & \mathbf{[T_n]}
\end{bmatrix}_{NKT \times 2K} \hspace{1mm},\hspace{10mm} X = \begin{bmatrix} \textbf{1}_{T_1} & \mathbf{[T_1]} & \textbf{1}_{T_1} & \textbf{0} & \textbf{0}
    \\ \vdots & \vdots & \textbf{0} & \ddots & \textbf{0} \\ \textbf{1}_{T_n} & \mathbf{[T_n]} & \textbf{0} & \textbf{0} & \textbf{1}_{T_n}
\end{bmatrix}_{NKT \times K+2}\hspace{1mm}.\]}

We describe the estimated relationships between the nodes via the ``association of evolution" and the ``evolution of association", previously described for bivariate models by \citet{gao_estimating_2017}. We extend these concepts to the case with an arbitrary number of nodes. The association of evolution ``summarizes how the evolution of [node j] is associated with the evolution of [node k]" \citep[p.4]{gao_estimating_2017}. In the statistical verbiage, this would be described as the correlation between random intercepts, which estimates the association between the nodes at time 0, and the correlation between random slopes, which estimates the association between the longitudinal trajectories of the nodes: 
\begin{align}
    \rho_{intercept}^{jk} = \frac{\sigma_{a_j a_k}}{\sqrt{\sigma_{a_j}^2}\sqrt{\sigma_{a_k}^2}} \textrm{ and } \rho_{slope}^{jk} = \frac{\sigma_{b_j b_k}}{\sqrt{\sigma_{b_j}^2}\sqrt{\sigma_{b_k}^2}} \text{.}
\end{align}
When $\sigma_{a_j}^2$ or $\sigma_{a_k}^2$ is zero (for intercept, or the corresponding slope values) we say $\rho_{\cdot}^{jk} = 0$. The evolution of association summarizes how the association between node j and node k evolves over time, and is given by the marginal correlation which is a function of time $t$: 
\begin{align}
    \rho_{marginal}(t)^{jk} = \frac{\sigma_{a_ja_k} + t \sigma_{a_jb_k} + t \sigma_{a_kb_j} + t^2 \sigma_{b_jb_k}}{\sqrt{\sigma_{a_j}^2 + 2t\sigma_{a_jb_j} + t^2\sigma_{b_j}^2 + \sigma_{\varepsilon_j}^2}\sqrt{\sigma_{a_k}^2 + 2t\sigma_{a_k b_k} + t^2\sigma_{b_k}^2 + \sigma_{\varepsilon_k}^2}} \text{.}
\end{align}
Advantages of the mixed-effects model include that the nodes do not need to be measured at the same time points, and that the number of repeated measurements does not need to be the same for all subjects or nodes. For simplicity and considering applications in -omics data, in describing our algorithms and in our simulations we assume without loss of generality that all nodes are measured at the same time points, such as $t \in \{1,2,...,T\}$, where $T$ is the number of repeated measurements. However, our algorithms are able to easily handle irregularly spaced timepoints.

\subsection{Algorithms} \label{sec:algorithms}

We propose two algorithms for estimating the fixed and random effects in the mixed-effects framework as described in equation (Eq. \ref{eq:mixeff}), i.e. the variance-covariance matrix for all pairs of nodes simultaneously. We refer to these algorithms as DCENt \ref{algo:1} and DCENt \ref{algo:2}, acronyms for Dynamic Co-Expression Network estimation over Time.

{\linespread{1.5}
\begin{algorithm}
    \caption{Estimation algorithm of DCENt 1, based on full $V$}
    \label{algo:1}
    \begin{algorithmic}
    \State \textbf{Output:} $\hat{D}$ and $\hat{E}$ are the estimated random-effect covariance and error variance, respectively
    \State \textbf{Require:} $K$ is number of nodes, $T$ is number of timepoints, $N$ is number of subjects, $Y$ is the data, vectorized in a nested order
    \State \textbf{Require:} $X,Z$ are the fixed- and random-effects design matrices (respectively). 
    \State \textbf{Initialization:} $\hat{D} \gets \frac{1}{K T} I_{K T}$
    \State \textbf{Initialization:} $\hat{\beta} \gets (X^\top  X)^{-1} X^\top  Y$
    \State \textbf{Initialization:} $\hat{r} \gets Y - X\hat\beta$
    \While{$\hat{\beta}$ and $\hat{V}$ are not converged} \Comment{estimating fixed effects and overall covariance $V$}
        \State $\hat{R} \gets \textrm{re-arranged}(\hat{r})$ \Comment{vector $r$ is re-arranged to be an N x KT matrix}
        \State $\hat{D}_0 \gets cov(\hat{R})$
        \State $\hat{V} \gets \textrm{diag}\left(\begin{bmatrix} 
          \hat{D}_0 & \dots & \hat{D}_0 \end{bmatrix}\right)$ \Comment{Block diagonal matrix with $\hat{D}_0$ repeated $N$ times}
        \State $\hat\beta \gets (X^\top  \hat{V}^{-1} X)^{-1} X^\top  \hat{V}^{-1} Y$
        \State $\hat{r} \gets Y - X\hat\beta$
    \EndWhile
    \State $\hat{V} \gets \textrm{threshold}(\hat{V})$ \Comment{adaptive thresholding of $\hat{V}$}
    
    \While{$\hat{D}$ and $\hat{\Omega}$ are not converged}
        \State $\hat{D} \gets (Z^\top  Z)^{-1} Z^\top  (\hat{V} - \hat{\Omega}) Z (Z^\top  Z)^{-1}$ \Comment{estimating $D$}
        \If{$\hat{D} \preceq 0$}
            \State $\hat{D} \gets \hat{D} + $ offset
        \EndIf
        \State $\hat{\Omega} \gets \textrm{diag}(\hat{V}) - \textrm{diag}(Z \hat{D} Z^\top )$ \Comment{estimating $E$}
    \EndWhile
    \State $\hat{D} \gets \textrm{threshold}(\hat{D})$ \Comment{hard threshold $D$ so it has the same proportion of 0s as $\hat{V}$}
    \For{$k \in \{1,...,K\}$}
        \State $\hat{E}^{(k,k)} \gets \textrm{mean}(\hat{\Omega}^{((k-1)T + 1,(k-1)T + 1)},...,\hat{\Omega}^{(kT,kT)})$ 
    \EndFor
\end{algorithmic}
\end{algorithm}
}

For DCENt 1 (algorithm \ref{algo:1}), we estimate the covariance using two distinct steps. First, we iterate between estimating the fixed effects coefficients $\boldsymbol\beta$ (fixed intercepts and slopes for all nodes) and the full covariance matrix $V = ZDZ^\top  + E$. We threshold $V$ so that it is sparse, which induces sparsity in $D$. Then, we solve for $D$ and $E$ from $\hat{V}$ (the estimator of V), again iterating between estimating the two values until the estimates $\hat{D}$ and $\hat{E}$ converge.

{\linespread{1.5}
\begin{algorithm}
    \caption{Estimation algorithm of DCENt 2, based on $D$ and $E$}
    \label{algo:2}
    \begin{algorithmic}
        \State \textbf{Output:} Same as Algorithm \ref{algo:1}
        \State \textbf{Require:} Same as Algorithm \ref{algo:1}
        \State \textbf{Initialization: } Same as Algorithm \ref{algo:1}
        \State \textbf{Initialization: } $\hat{\Lambda}_D \gets Cholesky(\hat{D} + I_{kt})$ 
        \While{$\hat{D}$ not converged}
            \State $\hat{V} \gets Z\hat{D}Z^\top  + \hat{E}$
            \State $\boldsymbol{\hat\beta} \gets (X^\top \hat{V}^{-1}X)^{-1} X^\top  \hat{V}^{-1}Y$ \Comment{estimate fixed effect coefficients}
            \State $\boldsymbol{\hat{b}} \gets \hat{\Lambda}_D (\hat{\Lambda}_D^\top Z^\top \hat{E}^{-T}\hat{E}^{-1}Z\hat{\Lambda}_D + I_{k})^{-1}$ \Comment{random effect coefficient conditional on the data}
            \State \hspace{8mm} $\times \hat{\Lambda}_D^\top Z^\top \hat{E}^{-T}\hat{E}^{-1}(Y - X\hat{\boldsymbol{\beta}})$
            \State $\boldsymbol{r_E} \gets {Y} - \boldsymbol{X\hat{\beta}}- Z\boldsymbol{\hat{b}}$. 
            \State $\hat{E}_{ii} \gets \textrm{var}(\hat{r}_{Ei})$ \Comment{error term variance}
            \State $ \hat{M} \coloneqq \hat{\Lambda}_D^{-1} (Z^\top Z)^{-1} Z^\top $ \Comment{temporary matrix for ease of notation}
            \State $\boldsymbol{\hat{\varepsilon}} \gets \hat{\Lambda}_{E_{nt}} (\hat{\Lambda}_{E_{nt}}^\top \hat{M}^\top \hat{M}\hat{\Lambda}_{E_{nt}} + I_{k})^{-1}$ \Comment{error term conditional on the data}
            \State \hspace{8mm} $\times \hat{\Lambda}_{E_{nt}}^\top \hat{M}^\top \hat{M}(Y - X\hat{\boldsymbol{\beta}})$
            \State $\boldsymbol{\hat{r}} \gets {Y} - X\boldsymbol{\hat{\beta}} - \boldsymbol{\hat\varepsilon}$. 
            \State $\boldsymbol{\hat{R}}_i \gets (Z_i^\top Z_i)^{-1}Z_i^\top  \boldsymbol{\hat{r}}_i$ \Comment{ith row of $\boldsymbol{\hat{R}}$ so $\boldsymbol{\hat{R}} \in \mathbb{R}^{n \times 2k}$}
            \State $\boldsymbol{\hat{D}} \gets threshold(\boldsymbol{\hat{R}})$ \Comment{adaptive thresholding of $\hat{R}$}
            \State $\hat{\Lambda}_{D} \gets Cholesky(\hat{\boldsymbol{D}} + I_{kt})$ \Comment{Cholesky decomposition of $\hat{D} + I_{kt}$}
        \EndWhile 
        \State $\hat\sigma^2 \gets \frac{1}{nkt} \|\hat{\Lambda}_V^{-1}(Y-X\boldsymbol{\hat{\beta}})\|^2$ \Comment{estimate of the variance kernel}
        \State $\hat{D} \gets \hat\sigma^2 \hat{\Lambda}_D \hat{\Lambda}_D^\top $ \Comment{adjust $D$ by the variance kernel}
        \State $\hat{E} \gets \hat\sigma^2 \hat{\Lambda}_E \hat{\Lambda}_E^\top $ \Comment{adjust $E$ by the variance kernel}
    \end{algorithmic}
\end{algorithm}
}

For DCENt 2 (algorithm \ref{algo:2}), we use a five-step expectation-maximization style procedure. We repeatedly estimate the fixed effects coefficients, random effects coefficients (conditional on the data), random effect covariance, realized error, and error variance, each conditional on the others. In order to sparsify the random effect covariance, we use the adaptive thresholding procedure as described in \citet{cao_large_2019}. For estimating the random effects, we use the maximum likelihood estimator for the random effect coefficient conditional on the data. See online Appendix A for complete details on the derivation of this estimator.

Both algorithms iterate between estimating the fixed effects coefficients and the random effects (RE) and error covariance. DCENt 1 estimates the entire marginal covariance ($V = ZDZ^\top  + E$ for random effect covariance matrix $D$ and error covariance matrix $E$) at each step, while DCENt 2 estimates the RE covariance matrix $D$ separately from the error covariance matrix $E$. DCENt 1 thus has the additional step of estimating $D$ and $E$ from the estimated $V$. DCENt 1 requires no distributional assumptions, whereas DCENt 2 relies on the normality assumption because it utilizes maximum likelihood estimators in multiple steps. DCENt 2 utilizes adaptive thresholding of the random effects covariance matrix $D$, while DCENt 1 relies on hard thresholding for the $D$ matrix (though adaptive thresholding is used on the overall covariance matrix $V$). If each subject is only measured at one timepoint ($T=1$) then DCENt 2 should be used, as the relation $ZDZ^\top  + E = V$ utilized by DCENt 1 is non-identifiable.

\subsection{Notes on theoretical guarantees} \label{subsec:theory}

In the Frobenius norm the thresholding operator has been shown to have optimal properties under certain regularity conditions. For example, when the input ($\mathbf{R}$) to the thresholding function has some tail-bounding properties (such as sub-Exponential tails) with covariance in the uniformity class of covariance matrices invariant under permutations: \[\mathcal{U}_\tau (q, c_0(K),M) = \left\{ \Sigma : \sigma_{ii} \leq M, \sum_{j=1}^K |\sigma_{ij}|^q \leq c_0(K) \textrm{ for all } i\right\} \textrm{ for } 0 \leq q \leq 1\] the Frobenius error of the thresholding estimator has been shown to follow \[O_P\left(K c_0(K) \left(\frac{\log K}{N}\right)^{1-q/2}\right),\] and a similar rate has been shown for a weighted class of covariance matrices \citep{bickel_covariance_2008, cai_adaptive_2011}. However, it is an open question whether our input to the thresholding estimator can be guaranteed to follow these regularity conditions, due to the fact that in both of our algorithms our input to the thresholding estimator is not the data itself but rather an estimation of the residuals for the intermediate model. 

\section{Simulation studies}

To evaluate these algorithms we perform comprehensive simulation studies. We generate simulated data using the set of mixed effects models (Eq. \ref{eq:mixeff}) directly (section \ref{sec:sim_data}) and using more general forms for the total covariance matrix $V$ which may not fit directly into the mixed effects framework, such as autoregressive and Toeplitz structures (online Appendix B). This allows us to evaluate both how the model performs when the assumed true model is correct as well as to establish a baseline when the model is mis-specified. For each of the models described below, and each combination of $N\in\{30, 50, 100\}$ subjects, $K \in \{5, 10, 20, 40, 100\}$ nodes, $T \in \{5,10\}$ timepoints, we generate the mean and variance parameters. From each set of parameters, we generate 10 datasets for testing. All simulations were generated with Julia version 1.10.2 \citep{Julia-2017} and the analysis was performed with R version 4.3.0 \citep{R-2023} on Northwestern's Quest system (Intel Xeon Gold 6338 CPUs running Red Hat Enterprise Linux 8). DCENt 1 step 1 is considered to have converged when the average percent change of the $\beta$ estimate and the $V$ estimate is less than $0.01\%$ ($\|\beta - \beta_{\textrm{prev}}\|_2 / \|\beta_{\textrm{prev}}\|_2 + \|V - V_{\textrm{prev}}\|_F / \|V_{\textrm{prev}}\|_F < 10^{-4}$) for two subsequent iterations. DCENt 1 step 2 is considered to have converged when the average percent change between the current and previous $D$ and $E$ estimates are below $10^{-4}$ for two subsequent iterations. For DCENt 2 the same threshold is used and $\Lambda_D$, $\Lambda_E$, and $\beta$ are considered. These thresholds were chosen arbitrarily based on preliminary runs, where the differences were seen to asymptote around these values. In each case, we compare the results of our algorithms to that of the pairwise method suggested by \citet{fieuws_pairwise_2006} based on $L_1$ error, Frobenius error, and the percentage of matrix entries estimated in the correct direction (positive, negative, or zero). While we would ideally compare to the method of \citet{ahn_momentbased_2012} as well, there is no publicly available code for this method.

\subsection{Simulations based on mixed-effects models} \label{sec:sim_data}

We begin by generating the data from the set of mixed effects models as described in Eq. (\ref{eq:mixeff}). We use two different settings for D, the random effect coefficient matrix: unstructured or compound symmetry. For the unstructured coefficient matrix, we don't impose any structure on the matrix. To generate these covariance matrices we draw
the diagonal entries as \(\sigma_{a_{j}} \sim U(2.5,3)\) and
\(\sigma_{b_{j}} \sim U(2,2.5)\), and the off-diagonals
as \(\sigma_{a_{j}b_{k}} \sim U(0.2,0.6) \times \text{ Rademacher}(0.5)\).

The compound symmetry (CS) covariance structure assumes constant
variance and equal correlation between all time points within a
variable, and constant cross-variable correlations. Our goal is for the overall covariance matrix ${V}$ to have the compound symmetry structure. We can achieve this by setting each row and column associated with the time-varying random effect to zero, setting each random slope variance to $\sigma_{a_i}^2$ (for node $i$), and setting the covariance between each pair of random intercepts to $\rho_{ij} \sigma_{a_i} \sigma_{a_j}$ (for nodes $i$ and $j$). 
To generate these matrices we draw $\sigma_{a_j} \sim U(2,3)$ and
$\rho_{ij} \sim U(0.2,0.8) \times \text{Rademacher}(0.5)$. For both settings, we draw the error variance values as \(\sigma_{k} \sim U(5,10)\). The outcome \(\mathbf{Y}_{i}\) for subject \(i\) is calculated directly according to the mixed effects model described above. The \(\beta\) and \(\alpha\) are drawn as \(\alpha_{j} \sim N(1,0.2)\) and \(\beta_{j} \sim N(0,0.2)\).

\subsection{Results}

Our main goal is the estimation of the $D$ matrix - the matrix describing the covariance of the random effects, so we begin there. Because we do not generate true $D$ matrices for the AR and Toeplitz models, we restrict this analysis to the compound symmetry and unstructured covariance matrix types. In practice, we find that both algorithms converge relatively quickly, in fewer than 10 iterations.

\subsubsection{Estimation of the Covariance Matrix of the Random Effects} 
\begin{figure}[h]
    \centering
    \begin{subfigure}{0.5\textwidth}
        \centering
        \includegraphics[width=2.5in,height=\textheight]{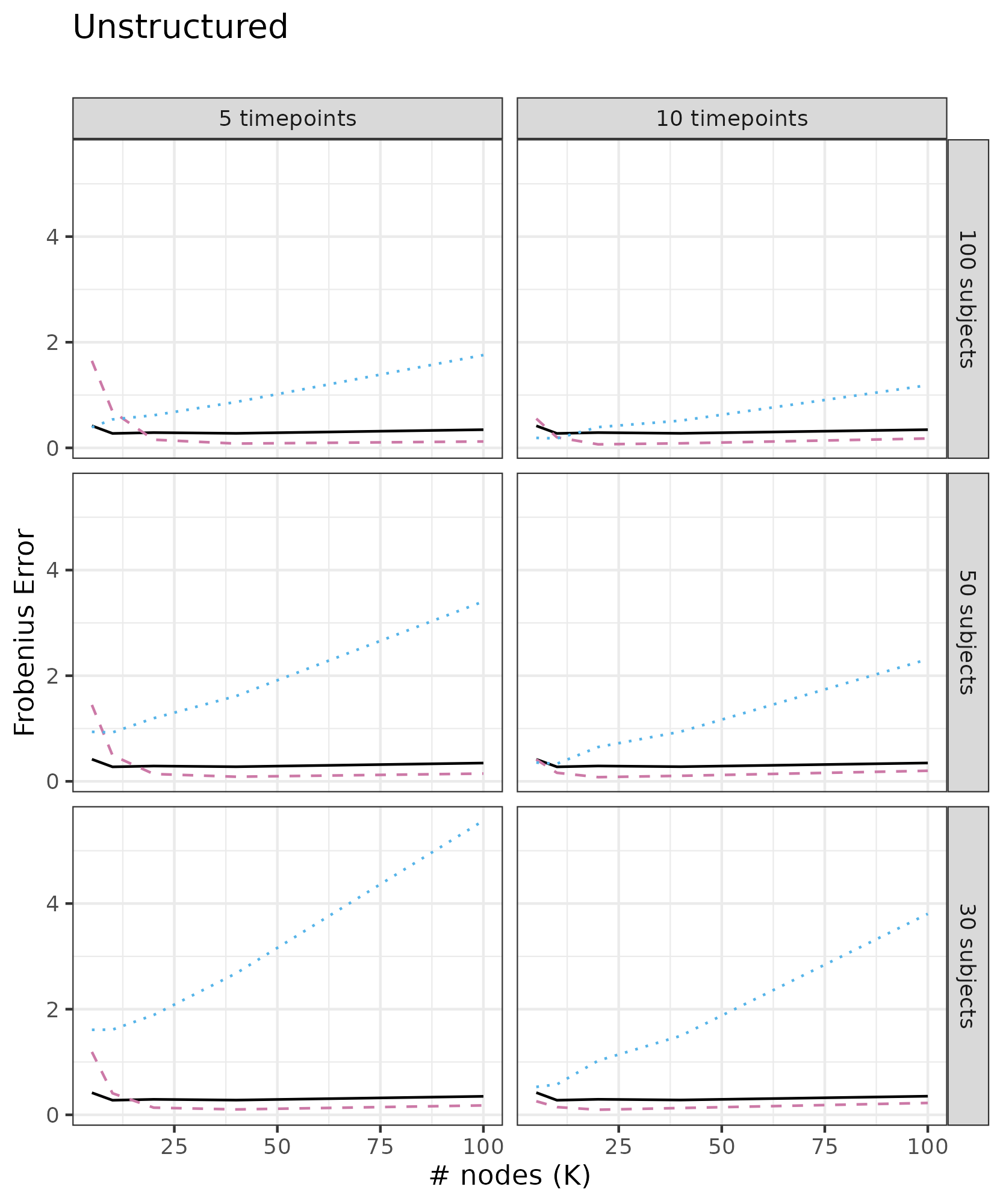}
    \end{subfigure}%
    \begin{subfigure}{0.5\textwidth}
        \centering
        \includegraphics[width=3in,height=\textheight]{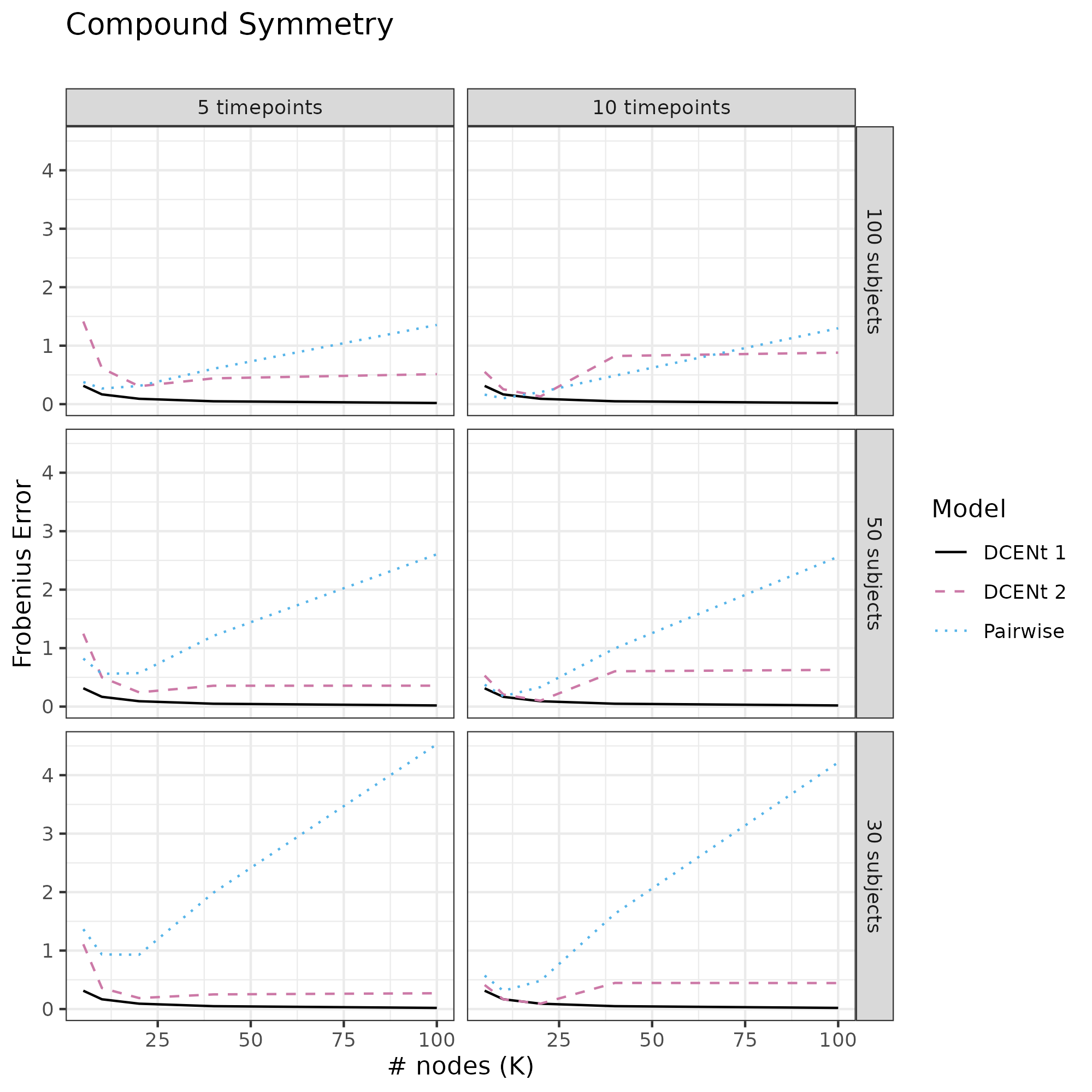}
    \end{subfigure}
    \caption{The Frobenius error in estimating the covariance of the random effect coefficients for each of the three algorithms. Panels are split between number of timepoints and number of subjects, increasing from bottom left to top right.}
    \label{fig:D_frob_1}
\end{figure}

\begin{figure}[h]
    \centering
    \begin{subfigure}{0.5\textwidth}
        \centering
        \includegraphics[width=2.5in,height=\textheight]{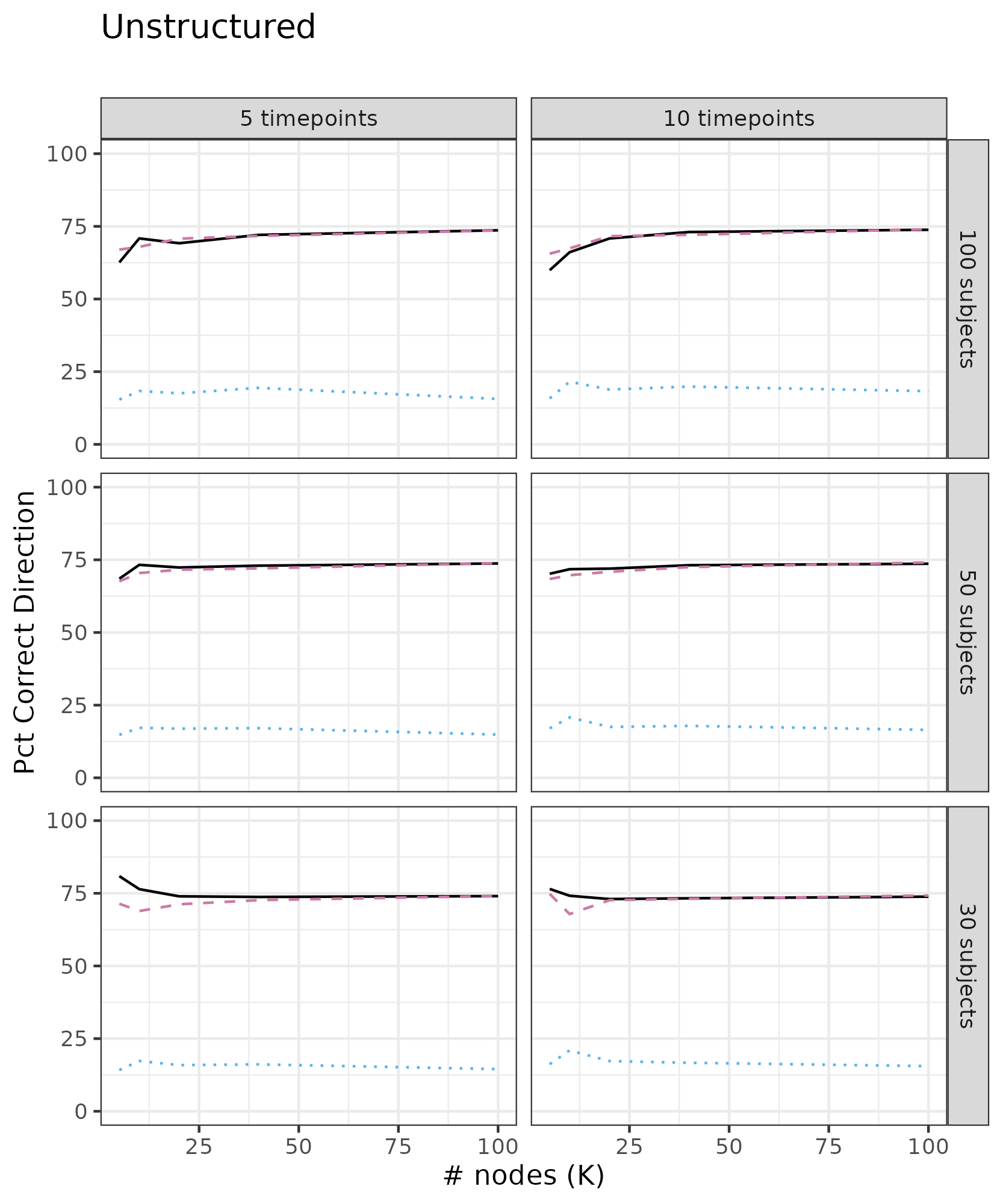}
    \end{subfigure}%
    \begin{subfigure}{0.5\textwidth}
        \centering
        \includegraphics[width=3in,height=\textheight]{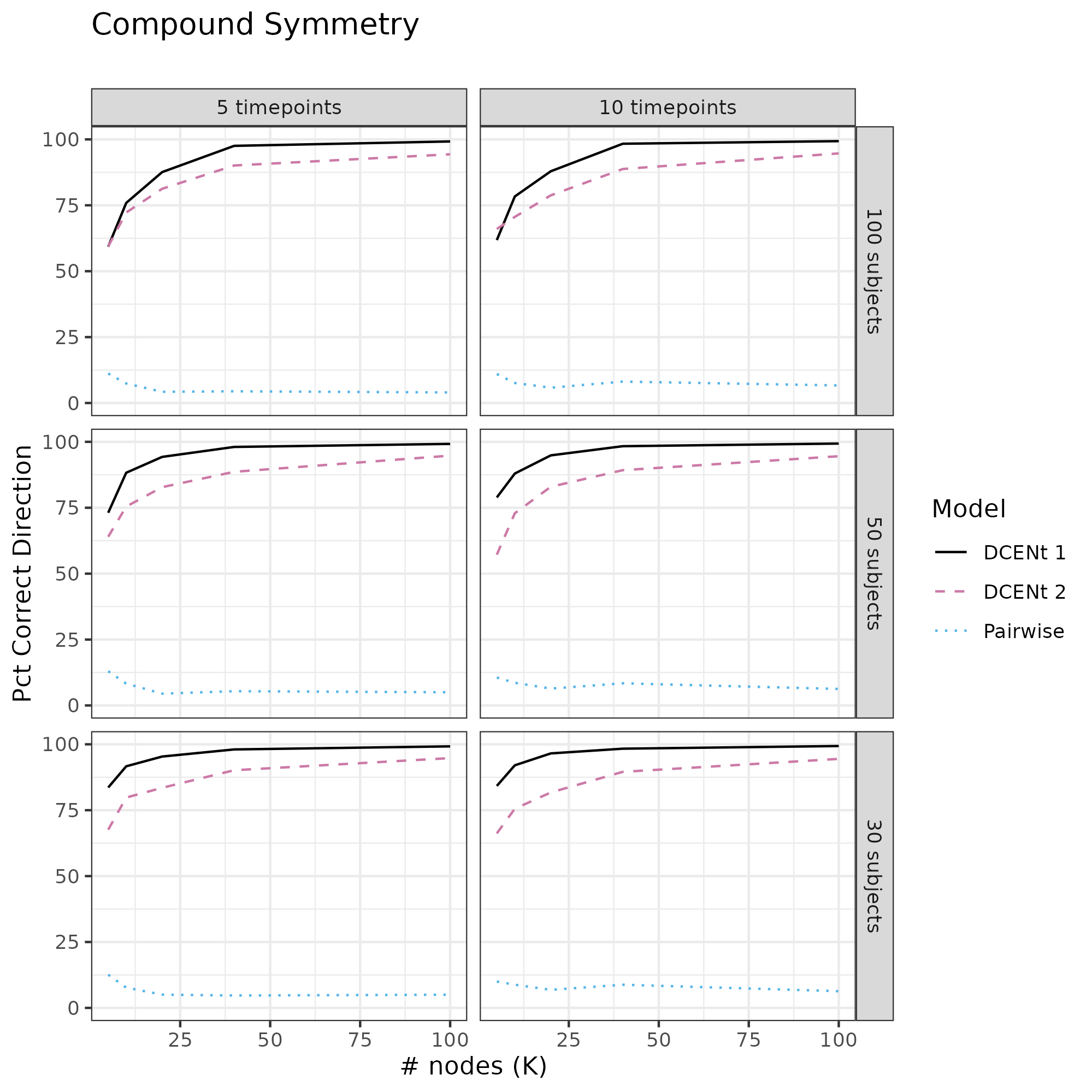}
    \end{subfigure}
    \caption{Percentage of pairwise relationships whose direction was correctly identified. Panels are split between number of timepoints and number of subjects, increasing from bottom left to top right.}
    \label{fig:D_posneg_1}
\end{figure}

We first examine the $D$ matrix, which is the covariance matrix of the random effect coefficients. In terms of Frobenius error, both algorithms outperform the pairwise estimation except potentially when there are very few nodes. This is especially true when the sample size is large (Figure \ref{fig:D_frob_1}, top panels). In terms of $L_1$ error,  all algorithms improve with the number of nodes. The two proposed DCENt algorithms perform similarly, both better than the pairwise approach (see online Appendix C). We also look at whether the algorithms can correctly identify the sign of the relationship between each pairs of nodes (Figure \ref{fig:D_posneg_1}). For the data with unstructured covariance matrix, both algorithms are able to identify approximately 75\% of the relationship directions, though at low sample size DCENt 1 appears to lose accuracy as the number of nodes increases. For the data with compound symmetry covariance matrix, DCENt 1 slightly outperforms DCENt 2, though as the number of nodes increases, each algorithm appears to be able to identify nearly 100\% of the relationship directions. This is a task that the pairwise estimation procedure is particularly ill-suited to, never exceeding 25\% in any case. 
 
\subsubsection{Estimation of the Overall Covariance}

\begin{figure}[h]
    \centering
    \begin{subfigure}{0.5\textwidth}
        \centering
        \includegraphics[width=2.5in,height=\textheight]{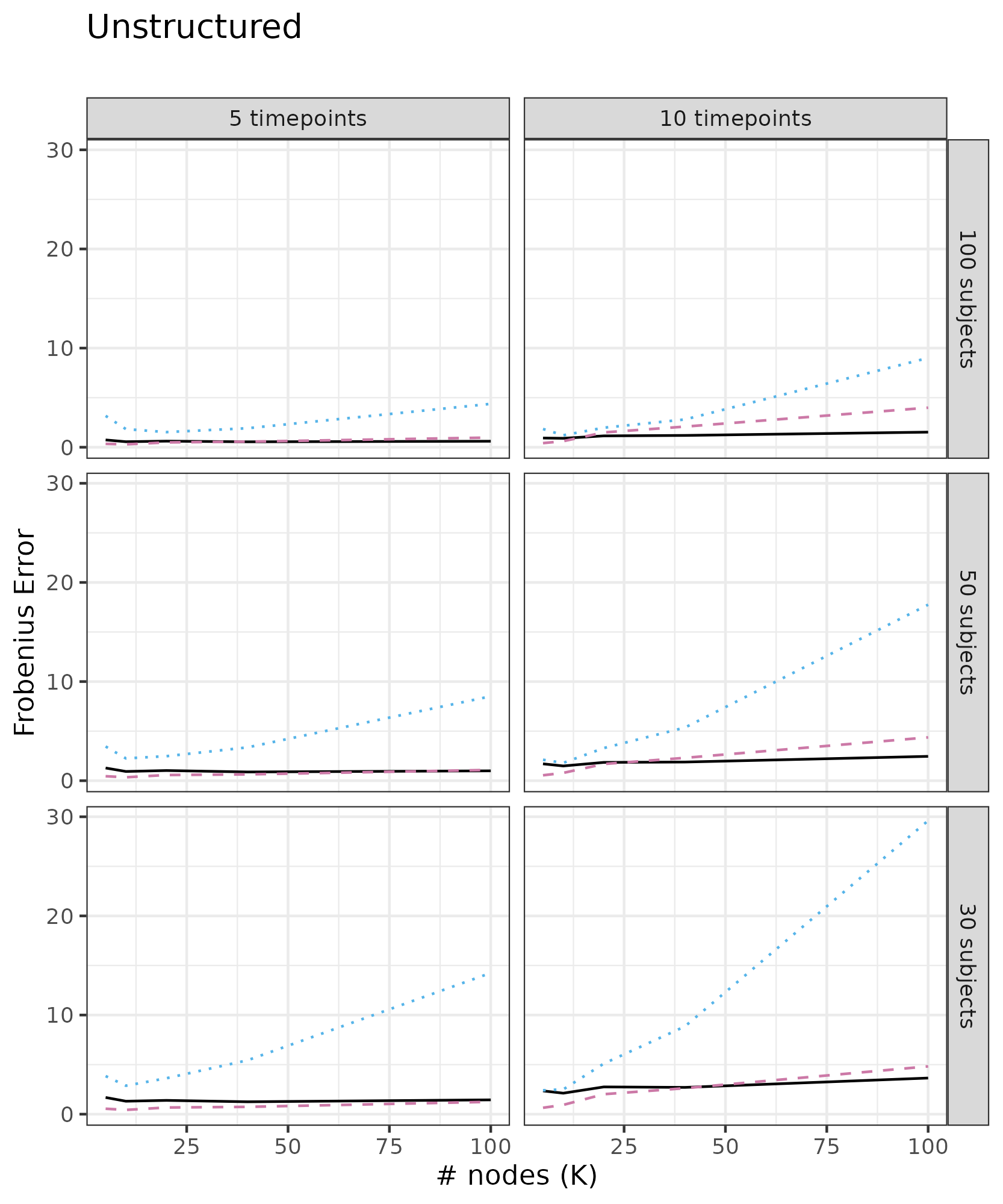}
    \end{subfigure}%
    \begin{subfigure}{0.5\textwidth}
        \centering
        \includegraphics[width=3in,height=\textheight]{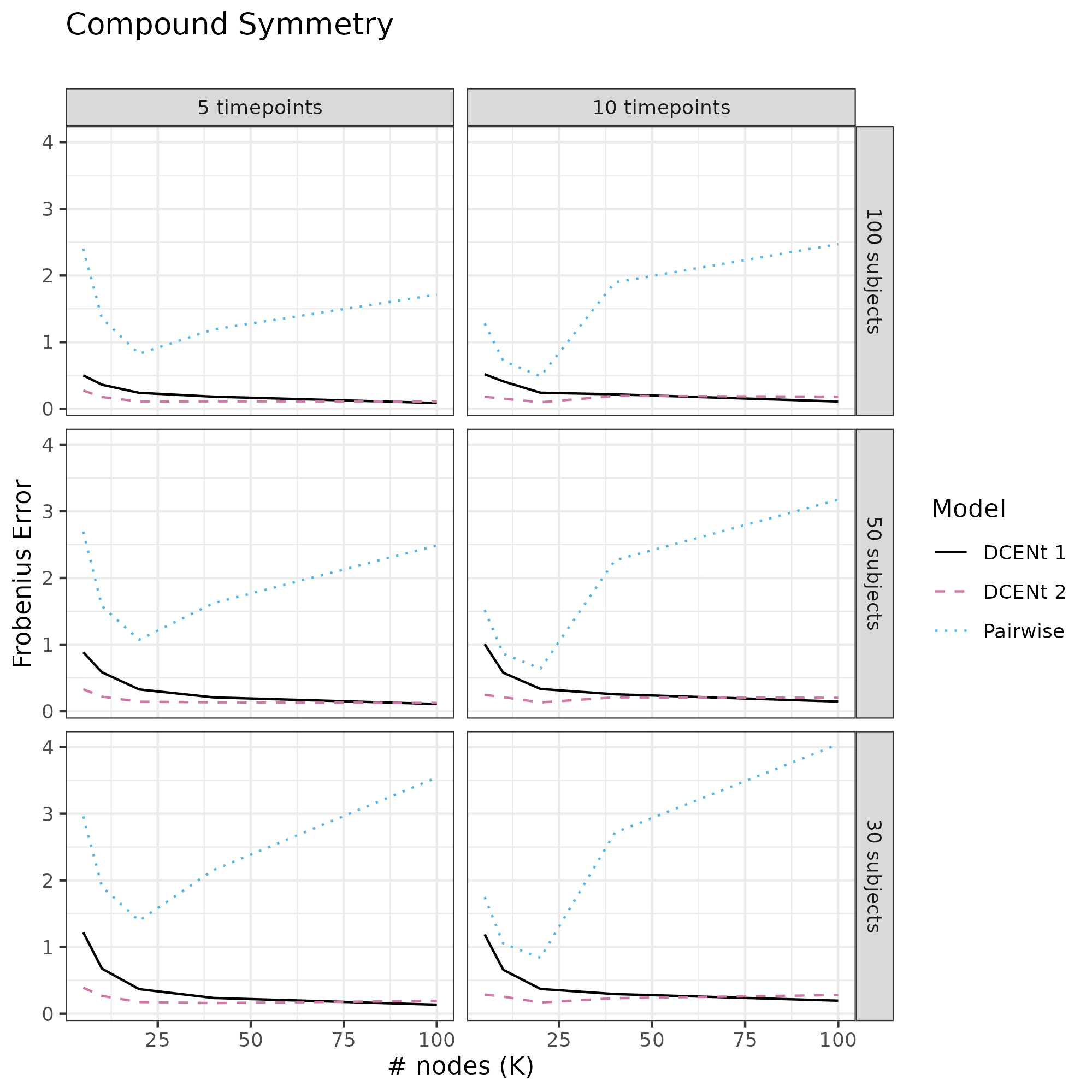}
    \end{subfigure}
    \caption{The Frobenius error in estimating the overall covariance for each of the three algorithms, unstructured and compound symmetric true covariance matrices. Panels are split between number of timepoints and number of subjects, increasing from bottom left to top right.}
    \label{fig:V_frob_1}
\end{figure}

Next we look at the overall covariance matrix, $V$. Here again we see both algorithms outperform the pairwise procedure in both the Frobenius (Figure \ref{fig:V_frob_1}) and the $L_1$ error (online Appendix C), for unstructured and compound symmetric covariance matrices. Again this is especially noticeable at low sample sizes and low number of timepoints. For the AR and Toeplitz matrices under model 1, DCENt 1 outperforms both DCENt 2 and the pairwise procedure in terms of Frobenius error, but the performance for the two algorithms is similar in terms of $L_1$ error (see online Appendix C). 

\subsubsection{Computational Time}

\begin{figure}[h]
    \centering
    \includegraphics[width=3.5in,height=\textheight]{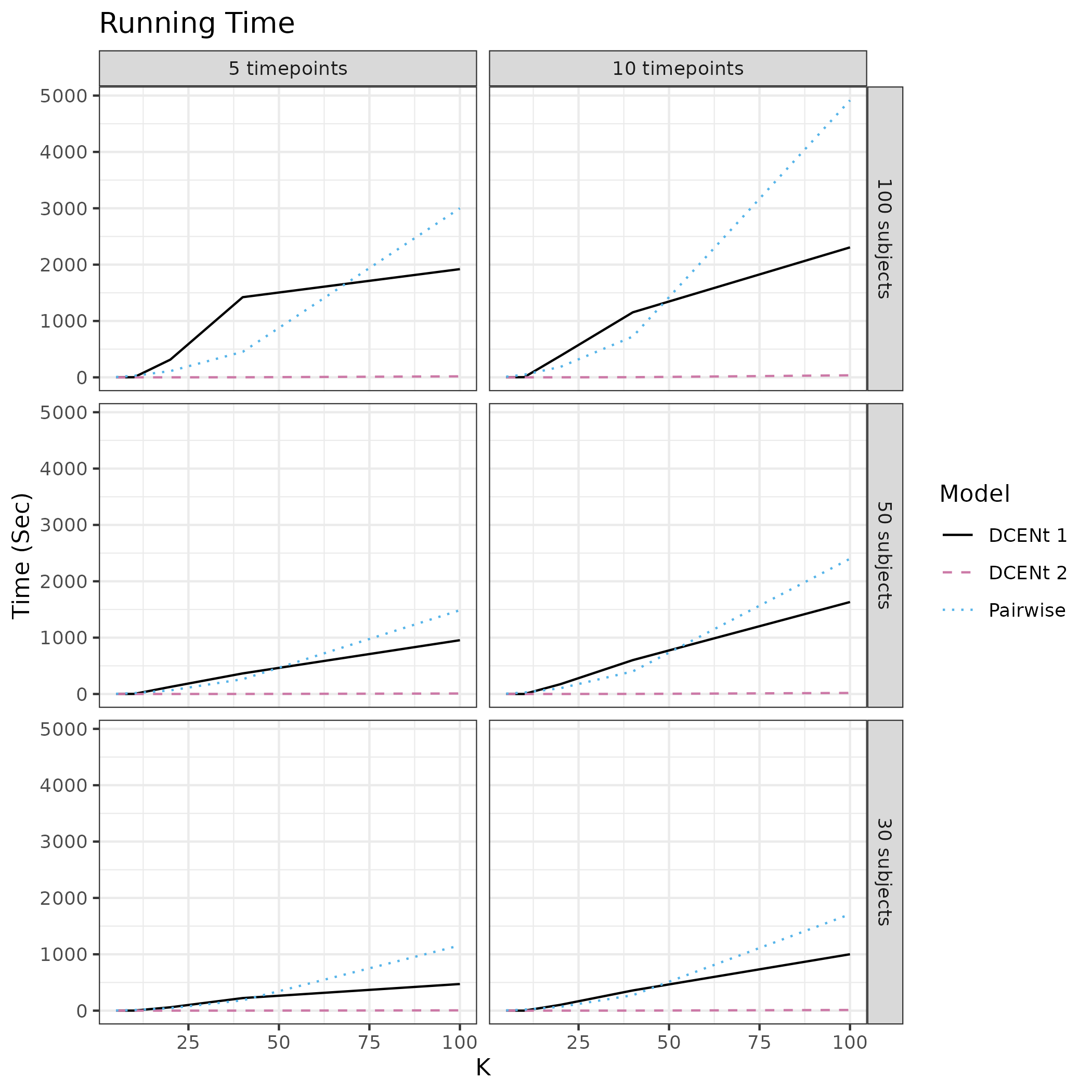}
    \caption{Average run time (in seconds) across all covariance types and data generation models. Panels are split between number of timepoints and number of subjects, increasing from bottom left to top right.}
    \label{fig:runtime}
\end{figure}

In terms of algorithm runtime, in general both algorithms are faster than the pairwise approach, especially when there are many nodes and many subjects (Figure \ref{fig:runtime}). This is at least partially due to the fact that the running time of the two algorithms is greatly reduced by taking advantage of the block structure of the covariance matrices (see online Appendix D for details). Both algorithms scale much better than the pairwise approach in terms of the number of subjects and number of nodes.

\section{Real Data Analysis}

We analyze the dynamic co-expression network based on participants in the CARDIA study. This dataset contains normalized protein abundances for 5346 proteins at 4 timepoints across 15 years (years 15, 20, 25, and 30 of the study) for 3123 subjects. Timepoints are t-normalized before beginning, and so take on values -1.6, -0.39, 0.39, and 1.16. We remove any subjects who do not have complete data for at least two timepoints, resulting in 2529 remaining. Because the number of proteins at each timepoint is so large and most of them are assumed to be unimportant, we work on a pre-specified set of proteins of interest. While computationally this is not necessary with our methods, it makes the results more interpretable and allows us to focus on proteins with more interesting trajectories. We selected a panel of 35 proteins previously implicated in cardiovascular disease and related outcomes, including inflammatory markers (IL6, TNF, IL1B, and CCL2), anti-inflammatory markers (IL10, IL19, IL20, IL22, and IL24), cardiac injury and stress markers (TNNI3, TNNT2, NT-proBNP, BNP/NPPB, and GDF15), lipid-related proteins (APOB, APOA1, LPA, and PCSK9), coagulation factors (FGA, FGG, SERPINE1, and VWF), markers of fibrosis and extracellular matrix remodeling (LGALS3, IL1RL1, MMP9, and TIMP1), endothelial and vascular function markers (VEGFA, ICAM1, VCAM1, and SELE), and emerging multi-system biomarkers (ADM, ANGPT2, PLAUR, FGF23, and IGFBP7). These proteins have been consistently associated with cardiovascular risk, disease progression, and mortality in large cohort studies and comprehensive proteomics and biomarker reviews \citep{hwang_circulating_1997,ridker_elevation_2000,whincup_willebrand_2002,yusuf_effect_2004,huang_comparative_2004,vaughan_pai1_2005,fibrinogenstudiescollaboration*_plasma_2005,braunwald_biomarkers_2008,ouyang_regulation_2011,lukasz_angiopoietin2_2013,mathew_fibroblast_2014,hodges_supar_2015,gandhi_insulinlike_2016,chow_role_2017,ridker_antiinflammatory_2017,sabatine_evolocumab_2017,ho_protein_2018a,braile_vegfa_2020,kronenberg_lipoproteina_2022,licordari_natriuretic_2024,ridker_inflammation_2024}. We fit the DCENt 1 and DCENt 2 algorithms to this data and find that the results are very similar between the two algorithms, so we focus on the results from DCENt 2. For means of comparison, we also look at the functional association network from FunCoup 6 based on protein expression data \citep{buzzao_funcoup_2025}. We do note, however, that these two networks are not expected to be exactly the same. While both are protein co-expression networks, the association of evolution network is based on the similarity of the trajectories of the proteins across time, while the FunCoup 6 network is a cross-sectional network based on the similarity of the expression levels of the proteins across subjects at one timepoint. Additionally, the edges represent different types of relationships: the FunCoup network is based on a Bayesian method and the edges represent probability of association. We consider an edge to be present in the association of evolution network if the absolute value of the correlation is greater than 0.3, the standard for fair strength of relationship established in \citet{akoglu_users_2018}. We consider an edge to be present in the FunCoup 6 network if the probability of association is greater than 0.5. Thus we take a somewhat liberal approach to defining edges in both networks. 

\begin{figure}
    \centering
    \includegraphics[width=2.75in,height=\textheight]{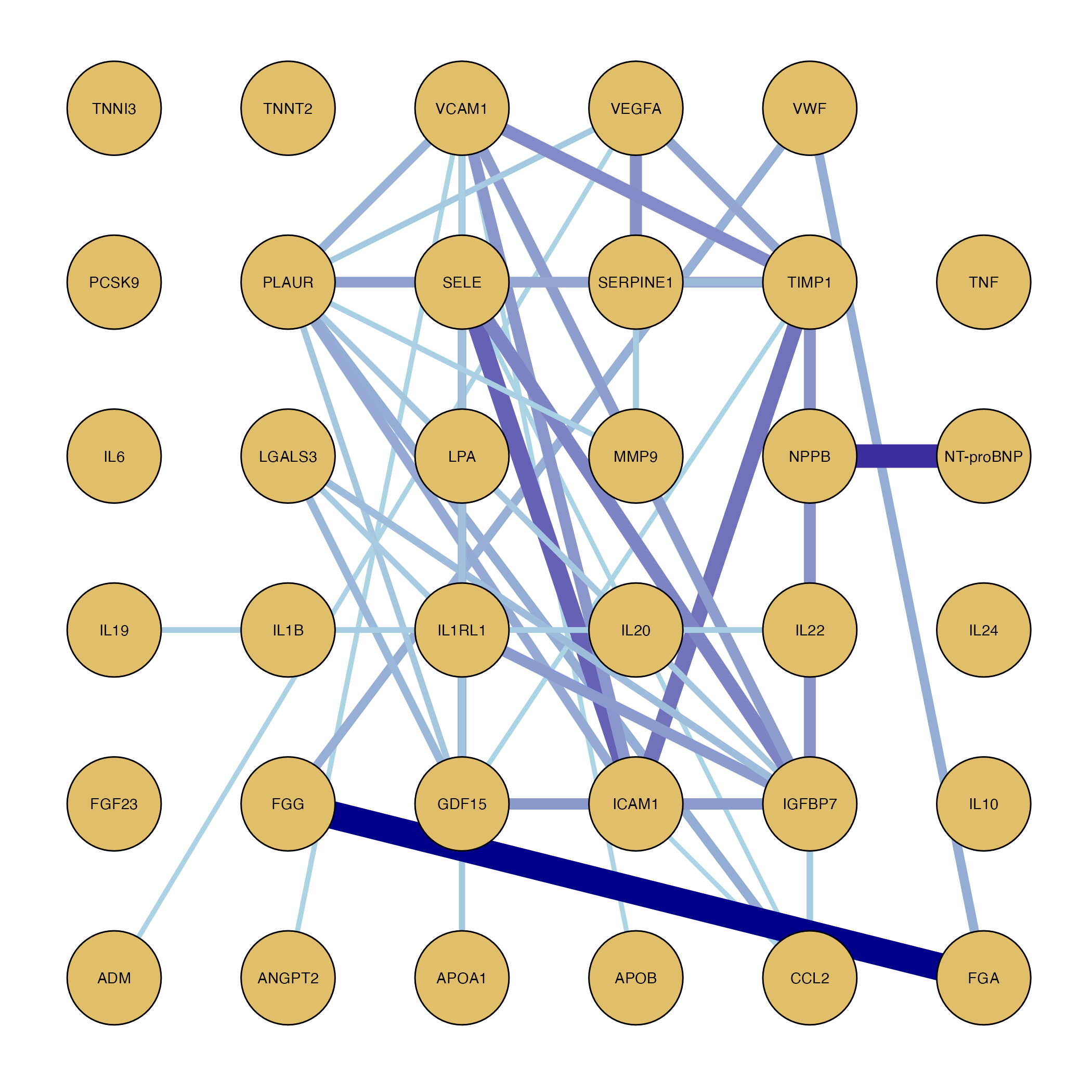}
    \includegraphics[width=2.75in,height=\textheight]{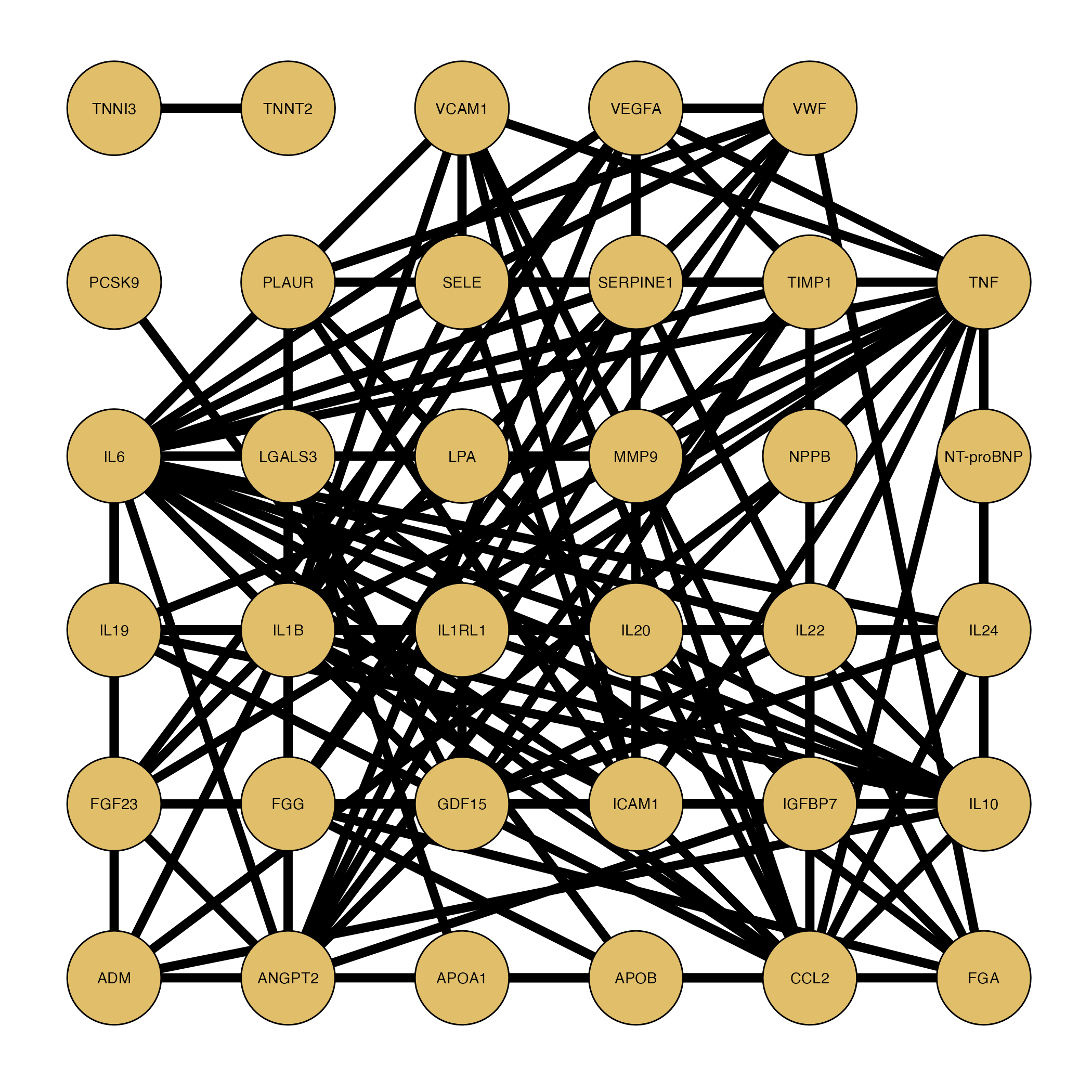}
    \caption{Left: Association of evolution for the slope, with edges colored and sized by the absolute value of the edge weight. Right: Functional association network from FunCoup 6 based on protein expression data. Note that the two networks are not expected to be the same, as they are based on different data and methods, and edges represent different types of evidence.}
    \label{fig:real_ae_slope}
\end{figure}

The association of evolution of the slopes (Figure \ref{fig:real_ae_slope}) is the greatest between Fibrinogen gamma chain (FGG; UniProt ID P02679) and Fibrinogen alpha chain (FGA; UniProt ID P02671), followed by Natriuretic peptides B (NPPB; UniProt ID P16860) and  N-terminal prohormone of brain natriuretic peptide (NT-proBNP). These relationships are not present in the FunCoup 6 network, which may be due to the fact that these proteins have similar trajectories across time but not necessarily similar expression levels across subjects at a given timepoint. As seen in Figure \ref{fig:real_ae_slope}, our association of evolution network does share many edges with the FunCoup 6 network. 

\begin{figure}
    \centering
    \includegraphics[width=3in,height=\textheight]{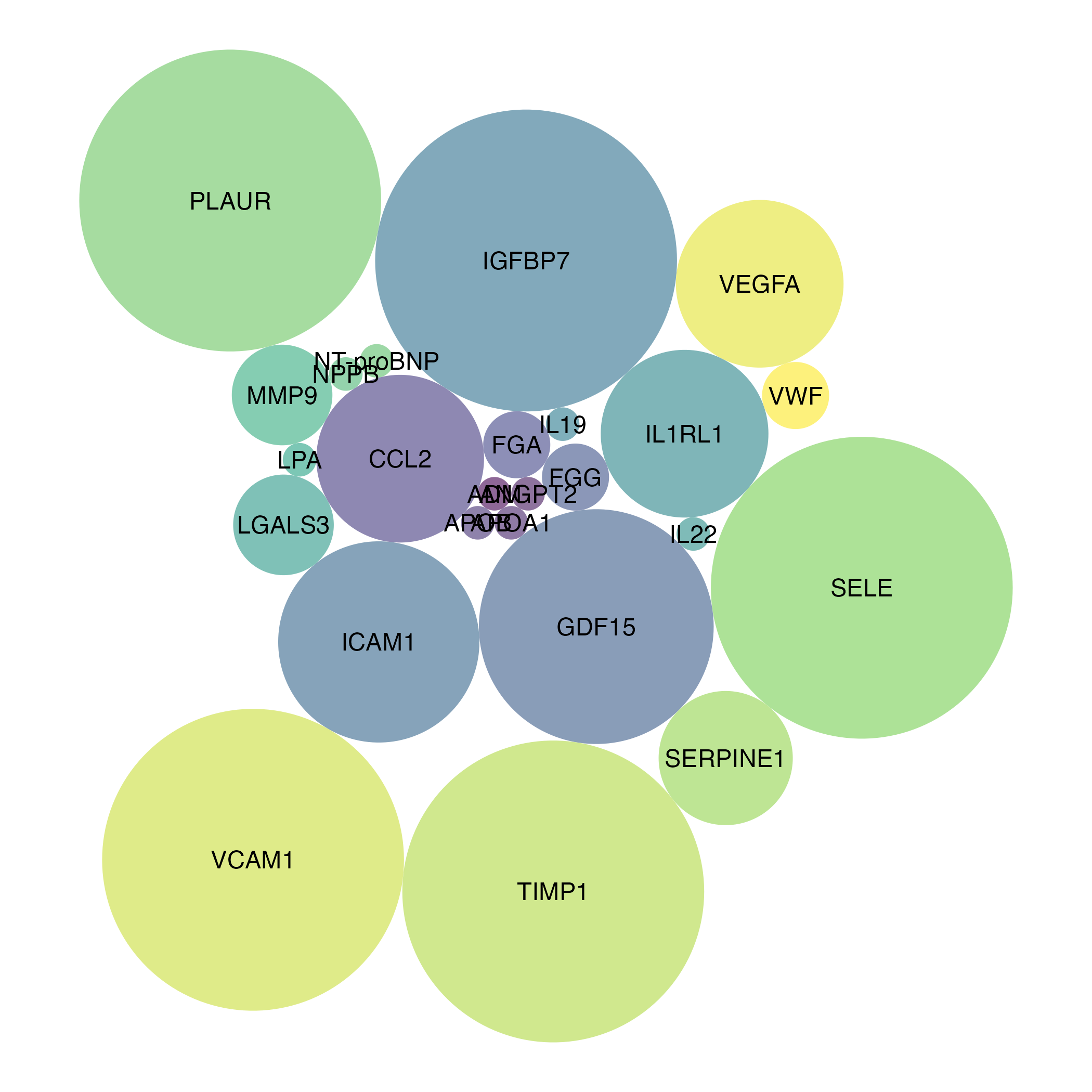}
    \includegraphics[width=3in,height=\textheight]{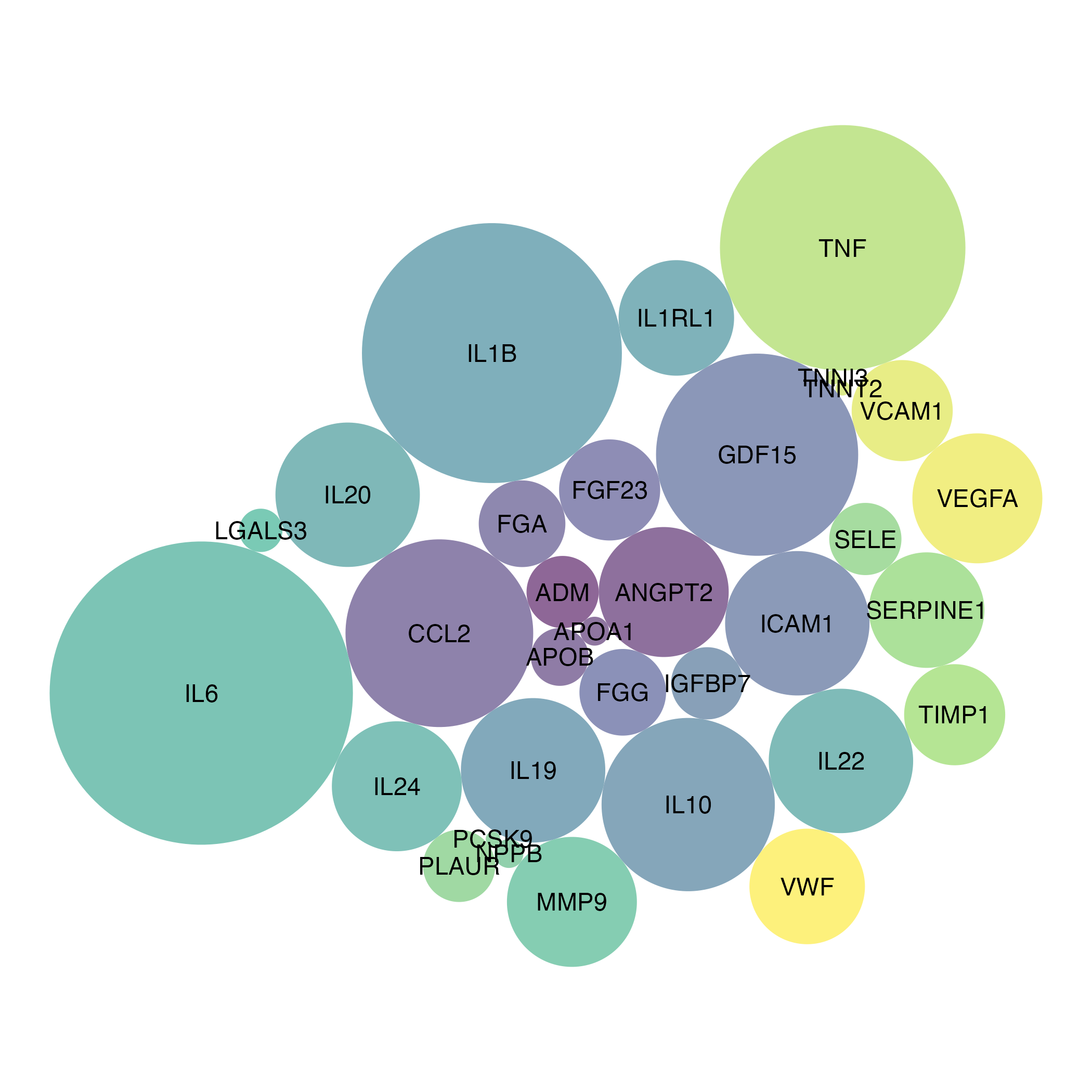}
    \caption{Node degree for association of evolution for the slope (left) and FunCoup 6 (right). Note that node colors here are not meaningful, and are only used to visually separate the nodes.}
    \label{fig:real_ae_slope_bubble}
\end{figure}

We also investigate the connectivity of the association of evolution network. Figure \ref{fig:real_ae_slope_bubble} shows each of proteins in the network, with bubbles sized by their degree (left: association of evolution, right: FunCoup 6). In the FunCoup 6 network, the major hub nodes are IL6 (P05231), TNF (P01375), IL1B (P01584), CCL2 (P13500), and GDF15 (Q99988). In our association of evolution network, the major hub nodes are IGFBP7 (Q16270), PLAUR (Q03405), SELE (P16581), TIMP1 (P01033), and VCAM1 (P19320). GDF15, and CCL2 have high degrees as well. Interestingly, the three most connected nodes in the FunCoup network (IL6, TNF, IL1B) are completely unconnected in our network. Also, all of our nodes with very strong correlation (FGG, FGA, NT-proBNP, NPPB) have low degree -- they are connected very strongly to each other, but not to much else. 

In terms of the mechanisms these proteins have been previously implicated in, the FunCoup network hub nodes are all inflammatory markers except GDF15 (cardiac injury and stress), whereas the association of evolution network hub nodes range across emerging multi-system biomarkers (IGFBP7, PLAUR), fibrosis and extracellular matrix remodelling (TIMP1), and endothelial and vascular function markers (SELE, VCAM1). Of our very strongly connected nodes, FGA and FGG are coagulation factors and NT-proBNP is a cardiac injury and stress marker.

\section{Conclusion and Future work}

In this paper, we employ a system of mixed-effects models to characterize dynamic co-expression networks between high-dimensional nodes arising from longitudinal data. Each node is specified through a mixed-effects model that explicitly accounts for between-subject heterogeneity as well as variability in temporal trajectories. Dependence among high-dimensional nodes is modeled via correlated random effects. The proposed method outputs variance-covariance matrices that capture dynamic co-expression patterns. Users can evaluate edge strength and stability to identify robust protein-protein associations and use node-level measures (e.g., degree or centrality) to detect hub proteins.

To this end, we introduce two thresholding-based algorithms for multivariate mixed-effects estimation with high-dimensional random effects methods. Via simulation studies, we show these methods outperform the pairwise approach, especially at small sample sizes. We also show that both of our algorithms are much faster than the pairwise approach. In DCENt 1 the overall covariance matrix is first fit and then decomposed into random effect and model error. In DCENt 2 the random effect error and model error matrices are optimized directly. 

Both algorithms are limited by the assumption of a linear relationship between the nodes and across time, though this is a common assumption in the literature. Furthermore, our modeling approach cannot handle the case where protein A at time t is correlated with protein B at time t+1, but not at time t. However, our approach has the advantage of allowing irregularly spaced timepoints and handling subjects with missing timepoints naturally. Additionally, both algorithms are much faster than the current standard approaches. 

Finally, we use both algorithms to analyze the dynamic protein co-expression networks of patients in the CARDIA study, and identify proteins with closely related co-expression patterns. In the future, this model should be extended to allow for variable selection of fixed effects, as well as random effects for which variable selection should not be applied. 

\section{Disclosure statement}\label{disclosure-statement}
The authors report there are no competing interests to declare.

\section{Data Availability Statement}\label{data-availability-statement}

Simulation data have been made available at the following url: \url{https://github.com/samozm/DCENt_supplement}. CARDIA data can be requested via \url{https://sites.uab.edu/cardia/}.

\section{Acknowledgements}\label{acknolwedgements}
The Coronary Artery Risk Development in Young Adults Study (CARDIA) is conducted and supported by the National Heart, Lung, and Blood Institute (NHLBI) in collaboration with the University of Alabama at Birmingham (75N92023D00002 \& 75N92023D00005), Northwestern University (75N92023D00004), University of Minnesota (75N92023D00006), and Kaiser Foundation Research Institute (75N92023D00003). This manuscript has been reviewed by CARDIA for scientific content.

\newpage
\bibliography{manuscript.bib}

\newpage
\appendix
\renewcommand{\thesection}{Appendix \Alph{section}}
\section{Conditional Random Effect Coefficient Estimates}\label{adx:coef_calc}

We want to estimate the random effects using the conditional mean (mean of effect conditioned on the data $y$). Here we introduce the kernel variable $u$ (what \citet{bates_fitting_2015a} refer to as the "spherical random effect") for estimating $\hat{b}$ and the kernel variable $v$ for estimating $\hat{\varepsilon}$. Using this variable allows us to essentialize the random parts of $y$ and $\mathcal{B}$ using the same variance parameter, which makes the calculations easier and allows the random effect matrices to be singular. These calculations follow loosely from \citet{bates_fitting_2015a,bates_mixedmodel_2025}. First we define the following: 

\begin{align*} 
    \Lambda_{E_{t}} \coloneqq \begin{bmatrix}
        \sigma_1 I_{k} \\ & \ddots \\ & & \sigma_k I_{k}
    \end{bmatrix} &
    \hspace*{10mm} \Lambda_{E_{nt}} \coloneqq \begin{bmatrix}
        \Lambda_{E_t} \\ & \ddots \\ & & \Lambda_{E_t}
    \end{bmatrix}
\end{align*}

\paragraph{Estimating $\varepsilon$ given $y$, $D$, and $E$}

\begin{align*}
    y - X\beta - \varepsilon_t &\sim N(0, ZDZ^T) \hspace{5mm}\textrm{ where } \varepsilon_t \coloneqq \begin{bmatrix}
        \varepsilon_1 \\ \vdots \\ \varepsilon_1 \\ \vdots \\ \varepsilon_k
    \end{bmatrix}
    \\ Z^T(y - X\beta - \varepsilon_t) &\sim N(Z^TZDZ^TZ)
    \\ (Z^TZ)^{-1} Z^T(y - X\beta - \varepsilon_t) &\sim N(0,D)
    \\ D &= \sigma^2 \Lambda_D\Lambda_D^T
    \\ \Lambda_D^{-1} (Z^TZ)^{-1} Z^T(y - X\beta - \varepsilon_t) &\sim N(0,I_{2k} \sigma^2)
    \\ \varepsilon_t &\sim N(0, \sigma^2 \Lambda_{E_{nt}} \Lambda_{E_{nt}}^T)
    \\ \textrm{Let }\varepsilon_t \eqqcolon \Lambda_{E_{nt}} v \textrm{ for }& v \sim N(0, \sigma^2 I_k) & 
    \\ \textrm{Let } M \coloneqq &\Lambda_D^{-1} (Z^TZ)^{-1} Z^T
    \\ f_{Y|V}(y|v) &= \frac{1}{(2\pi\sigma^2)^{2k/2}} \expfunc{-\frac{1}{2\sigma^2} \|M(y - X\beta - \Lambda_{E_{nt}} v)\|^2} 
    \\ f_V(v) &= \frac{1}{(2\pi\sigma^2)^{ntk/2}} \expfunc{-\frac{1}{2\sigma^2}\|v\|^2}
    \\ f_{V|Y}(v|y) &\propto \expfunc{-\frac{1}{2\sigma^2} \left(\|M(y - X\beta - \Lambda_{E_{nt}} v)\|^2 + \|v\|^2\right)} 
    \\ \textrm{Let } Q_M(v) &\coloneqq \|M(y - X\beta - \Lambda_{E_{nt}} v)\|^2 + \|v\|^2
    \\ \hat{v} = \mathbb{E}[V | Y=y] &= \underset{v}{\argmin} Q_M(v)
    \\ \frac{\partial}{\partial v} Q_m(v) &= -\Lambda_{E_{nt}}^TM^TM(y - X\beta -\Lambda_{E_{nt}} v) + v
    \\ &\hspace{60mm}\coloneqq 0
    \\ (\Lambda_{E_{nt}}^TM^TM\Lambda_{E_{nt}} + I_{nkt}) v &= \Lambda_{E_{nt}}^TM^TM(y - X\beta)
    \\ \hat{v} &= (\Lambda_{E_{nt}}^TM^TM\Lambda_{E_{nt}}+ I_{nkt})^{-1}\Lambda_{E_{nt}}^TM^TM(y - X\beta)
    \\ \hat{\varepsilon} &= \Lambda_{E_{nt}} (\Lambda_{E_{nt}}^TM^TM\Lambda_{E_{nt}} + I_{nkt})^{-1}\Lambda_{E_{nt}}^TM^TM(y - X\beta)
\end{align*}

\paragraph{Estimating $b$ given $y$, $D$, and $E$}

\begin{align*}
    y | \mathcal{B} = b &\sim N(X\beta + Zb, E_{nt}) 
    \\ (y - X\beta - Zb) &\sim N(0, E_{nt})
    \\ \textrm{Let } E_{nt} &\eqqcolon \sigma^2 \Lambda_{E_{nt}} \Lambda_{E_{nt}}^T
    \\ \Lambda_{E_{nt}}^{-1}(y - X\beta - Zb) &\sim N(0, I_{nkt})
    \\ \mathcal{B} &\sim N(0, \sigma^2 \Lambda_D \Lambda_D^T)
    \\ \textrm{Let } b \eqqcolon \Lambda_D u \textrm{ for }& u \sim N(0, \sigma^2 I_{2k})
    \\ f_{Y | U}(y|u) &= \frac{1}{(2\pi\sigma^2)^{nkt/2}} \expfunc{-\frac{1}{2\sigma^2} \|\Lambda_{E_{nt}}^{-1}(y - X\beta - Z\Lambda_D u) \|^2}
    \\ f_U(u) &= \frac{1}{(2\pi\sigma^2)^{2k}} \expfunc{-\frac{1}{2\sigma^2} \|u\|^2}
    \\ f_{U | Y}(u|y) & \propto \expfunc{-\frac{1}{2\sigma^2} \left(\|\Lambda_{E_{nt}}^{-1}(y - X\beta - Z\Lambda_D u) \|^2 + \|u\|^2\right)}
    \\ Q_E(u) &\coloneqq \|\Lambda_{E_{nt}}^{-1}(y - X\beta - Z\Lambda_D u) \|^2 + \|u\|^2
    \\ \hat{u} &= \underset{u}{\argmin} Q_E(u) 
    \\ \hat{u} &= (\Lambda_D^TZ^T\Lambda_{E_{nt}}^{-T}\Lambda_{E_{nt}}^{-1}Z\Lambda_D + I_{2k})^{-1}\Lambda_D^TZ^T\Lambda_{E_{nt}}^{-T}\Lambda_{E_{nt}}^{-1}(y - X\beta)
    \\ \hat{b} &= \Lambda_D (\Lambda_D^TZ^T\Lambda_{E_{nt}}^{-T}\Lambda_{E_{nt}}^{-1}Z\Lambda_D + I_{2k})^{-1}\Lambda_D^TZ^T\Lambda_{E_{nt}}^{-T}\Lambda_{E_{nt}}^{-1}(y - X\beta)
\end{align*}

\paragraph{Estimating $\sigma^2$}

\begin{align*}
    \\ V^\star &\coloneqq ZDZ^T + E_{nt}
    \\ &= Z(\sigma^2\Lambda_D \Lambda_D^T)Z^T + \sigma^2\Lambda_E\Lambda_E^T
    \\ V^\star &\eqqcolon \sigma^2 V
    \\ V &\coloneqq Z\Lambda_D\Lambda_D^T Z^T + \Lambda_E\Lambda_E^T
    \\ Y &\sim N(X\beta, \sigma^2V)
    \\ f_Y(y) &= \frac{1}{\sqrt{2\pi}{\sigma^2}^{nkt/2}|V|^{-1/2}} \expfunc{-\frac{1}{2\sigma^2}(y - X\beta)^TV^{-1}(y - X\beta)}
    \\ \ell(\sigma^2|V) &\propto -\frac{nkt}{2}\log \sigma^2 -\frac{1}{2\sigma^2} \|\Lambda_V^{-1}(y-X\beta)\|^2
    \\ \frac{\partial}{\partial \sigma^2} \ell(\sigma^2|V) &= -\frac{nkt}{2\sigma^2} + \frac{1}{2(\sigma^2)^2} \|\Lambda_V^{-1}(y-X\beta)\|^2
    \\ &\hspace{40mm} \coloneqq 0
    \\ \frac{nkt}{2\sigma^2} &=  \frac{1}{2(\sigma^2)^2} \|\Lambda_V^{-1}(y-X\beta)\|^2
    \\ \hat\sigma^2 &= \frac{1}{nkt} \|\Lambda_V^{-1}(y-X\beta)\|^2
\end{align*}

\section{Simulation $L_1$ Error Results}\label{adx:l1}

\FloatBarrier
\begin{figure}[h]
    \centering
    \begin{subfigure}{0.5\textwidth}
        \centering
        \includegraphics[width=2.5in,height=\textheight]{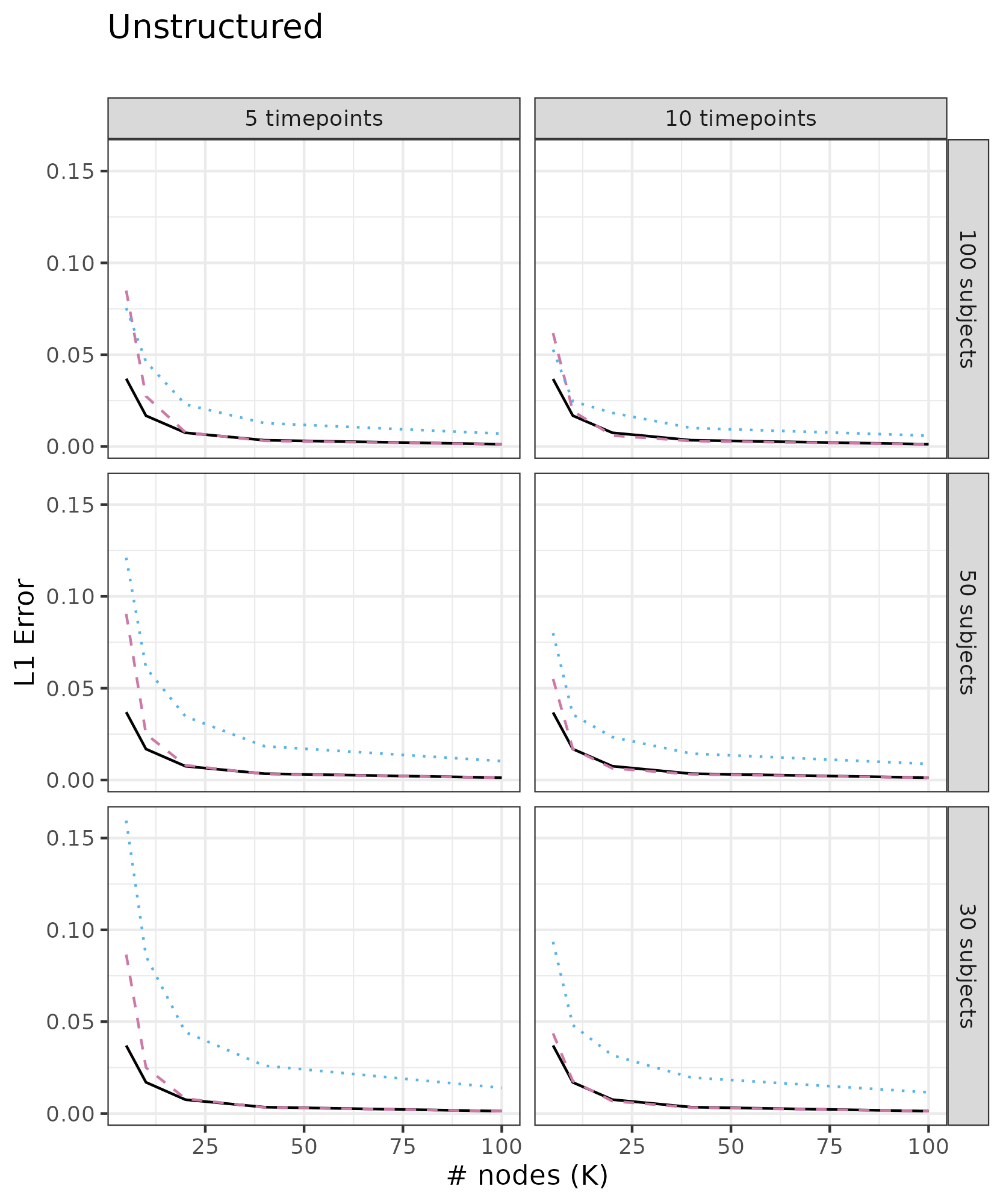}
    \end{subfigure}%
    \begin{subfigure}{0.5\textwidth}
        \centering
        \includegraphics[width=3in,height=\textheight]{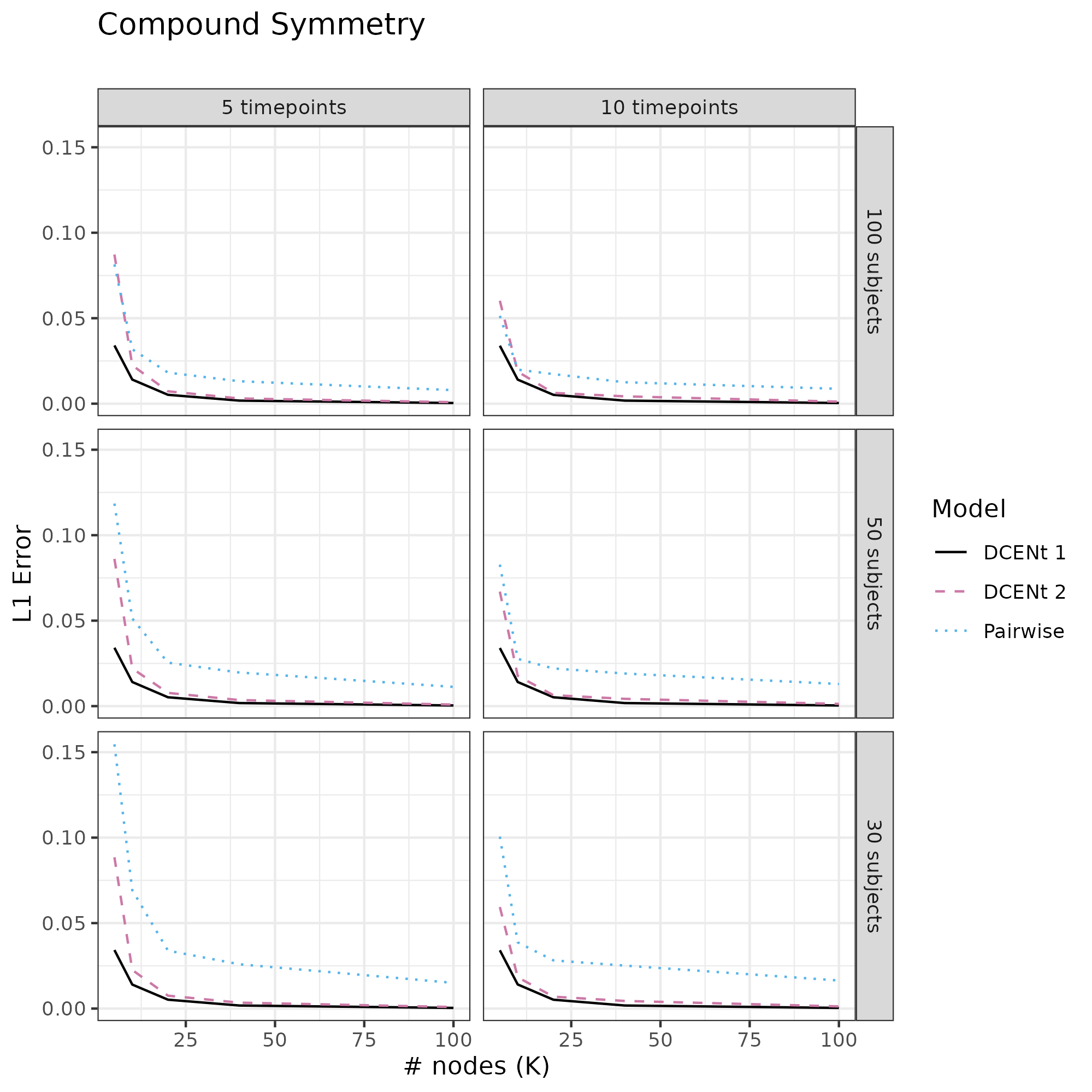}
    \end{subfigure}
    \caption{The $L_1$ error in estimating the covariance of the random effect coefficients for each of the three algorithms. Panels are split between number of timepoints and number of subjects, increasing from bottom left to top right.}
    \label{fig:D_l1_1}
\end{figure}

\begin{figure}[h]
    \centering
    \begin{subfigure}{0.5\textwidth}
        \centering
        \includegraphics[width=2.5in,height=\textheight]{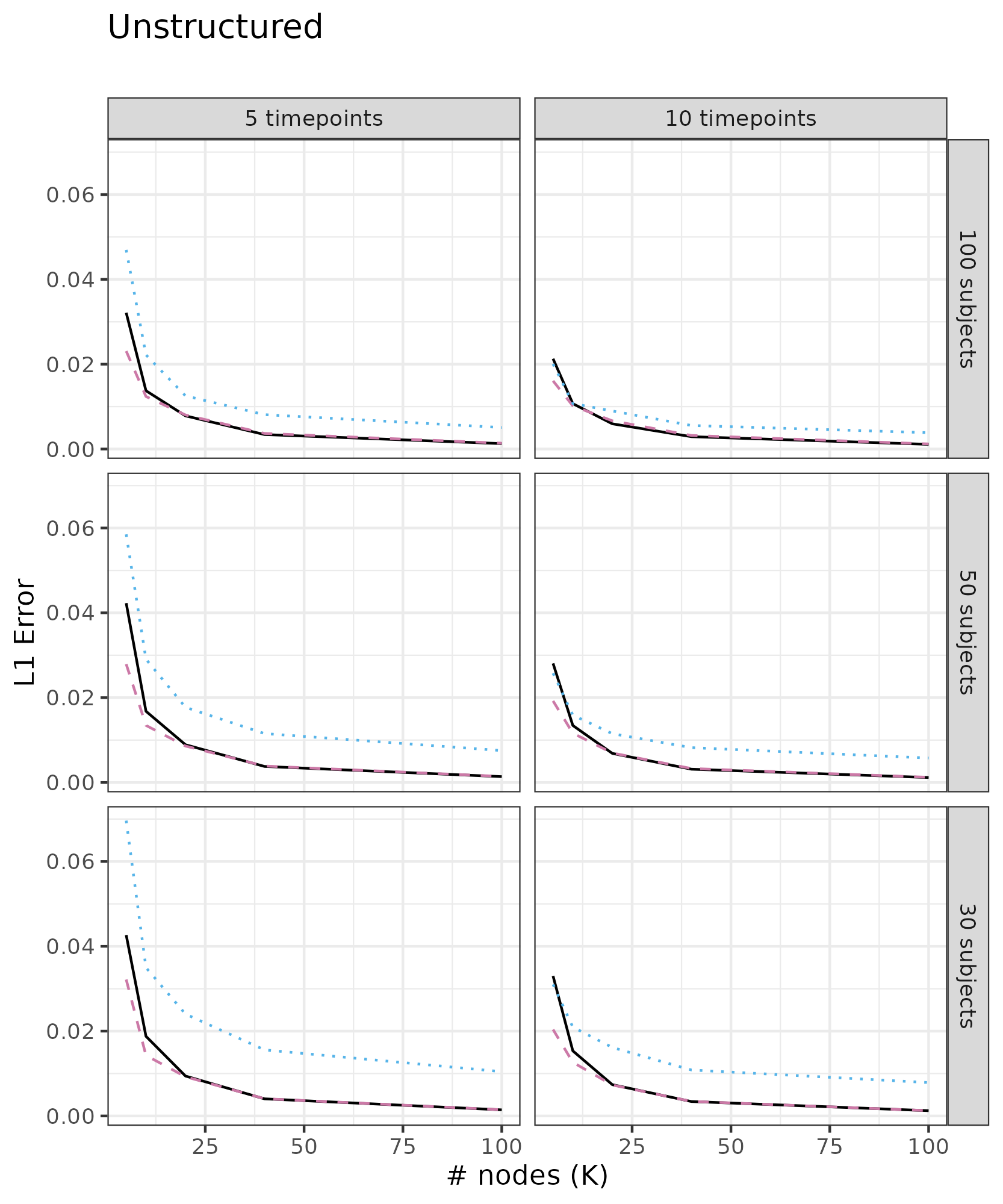}
    \end{subfigure}%
    \begin{subfigure}{0.5\textwidth}
        \centering
        \includegraphics[width=3in,height=\textheight]{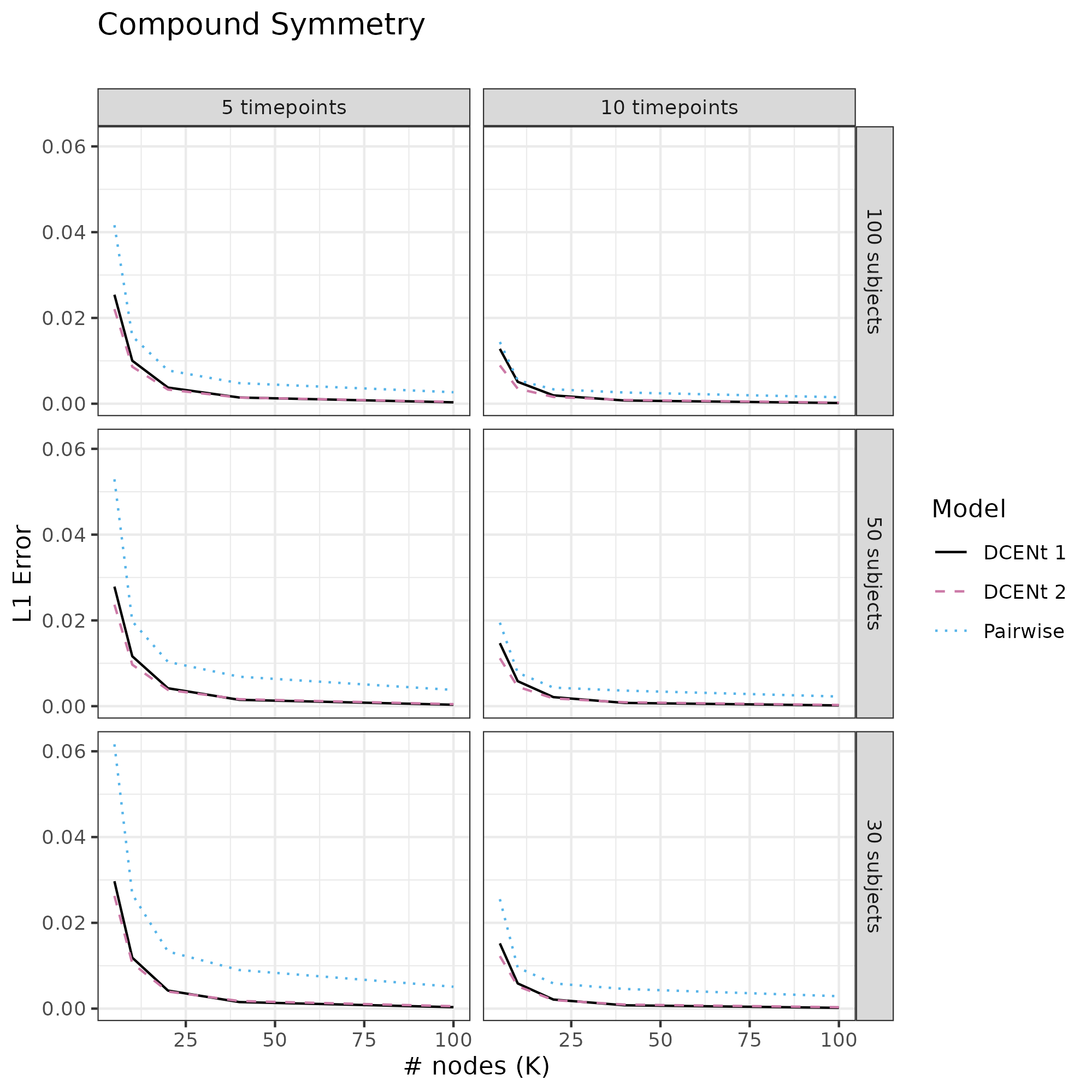}
    \end{subfigure}
    \caption{The $L_1$ error in estimating the overall covariance for each of the three algorithms, unstructured and compound symmetric true covariance matrices. Panels are split between number of timepoints and number of subjects, increasing from bottom left to top right.}
    \label{fig:V_l1_1}
\end{figure}
\section{Additional Simulations}\label{adx:adtl_sims}

\subsection{Simulations based on general covariance structures}

Following \citet{cao_large_2019}, the outcome for subject \(i\) is given by one of two models:

\begin{enumerate}
\item
  \(\mathbf{Y}_{i} \sim N\left( \mu_{i},\mathbf{R}_{i} \right)\)
\item
  \(\mathbf{Y}_{i} = \mathbf{\mu} + \frac{F_{i}U_{k}}{\sqrt{10}}\) where
  \(F_{i}F_{i}^{T} = R_{i}\) and the components of \(U_{k}\) are drawn from \(\Gamma(10,1)\)
\end{enumerate}
In the covariance matrix $K^\star$ is the number of connected nodes, chosen to impose sparsity. Without loss of generality, we choose the first $K^\star$ nodes to be connected, and choose $K^\star = \left[\frac{K}{2}\right]$. We draw the mean for each subject as $\mu_i \sim U(0,10)$. We use three different settings for the general covariance matrices: autoregressive, Toeplitz, and unstructured. For the unstructured covariance matrix, we draw the diagonals as $\sigma_i^2 \sim U(0.8,1.2)$ and the off-diagonals as $\sigma_{ij} \sim U(0.2,0.8) \times \text{Rademacher}(0.5)$.

The covariance matrix \(R\) looks like: 
{\renewcommand*{\arraystretch}{0.75}
\[R = \begin{pmatrix}
\Sigma_{1} & \Sigma_{12} & \ldots & \Sigma_{1K^\star} & \textbf{0} & ... & \textbf{0} \\
 & \Sigma_{2} & \ldots & \Sigma_{2K^\star} & \textbf{0} & ... & \textbf{0}\\
 & & \ddots & & \textbf{0} & ... & \textbf{0}\\
 & & & \Sigma_{K^\star} & \textbf{0} & ... & \textbf{0} \\
 & & & & \Sigma_{K^\star + 1} & \textbf{0} & \textbf{0}\\ 
 & & & & & \ddots & \textbf{0} \\
 & & & & & & \Sigma_{K}
\end{pmatrix}\textbf{.}\]}

\subsubsection{Autoregressive}

The autoregressive (AR(1)) covariance structure assumes correlations
decay exponentially with time lag within variables and between
variables. The within-variable covariance matrix is
{\renewcommand*{\arraystretch}{0.75}
\[\Sigma_{v} = \begin{pmatrix}
\sigma_{v}^{2} & \rho_{v}\sigma_{v} & \rho_{v}^{2}\sigma_{v} & \ldots & \rho^{T - 1}\sigma_{v} \\
 & \sigma_{v}^{2} & \rho_{v}\sigma_{v} & \ldots & \rho_{v}^{T - 2}\sigma_{v} \\
 & & \sigma_{v}^{2} & \ldots & \rho_{v}^{T - 3}\sigma_{v}^{2} \\
 \vdots & \vdots & \vdots & \ddots & \vdots \\
 & & & & \sigma_{v}^{2}
\end{pmatrix}\textbf{.}\]}

The between-variable covariance matrix is
{\renewcommand*{\arraystretch}{0.75}
\[\Sigma_{vw} = \begin{pmatrix}
\rho_{vw}\sigma_{v}\sigma_{w} & \rho_{v}\rho_{w}\rho_{vw}\sigma_{v}\sigma_{w} & \ldots & \rho_{v}^{T - 1}\rho_{w}^{T - 1}\rho_{vw}\sigma_{v}\sigma_{w} \\
 & \rho_{vw}\sigma_{v}\sigma_{w} & \ldots & \rho_{v}^{T - 2}\rho_{w}^{T - 2}\rho_{vw}\sigma_{v}\sigma_{w} \\
 & & \ddots & \vdots \\
 & & & \rho_{vw}\sigma_{v}\sigma_{w}
\end{pmatrix}\textbf{.}\]}

We draw $\sigma_v \sim U(0.8,1.2)$, $\rho_{v} \sim U(0.2,0.8) \times \text{Rademacher}(0.5)$, and $\rho_{vw} \sim U(0.2,0.8) \times \text{Rademacher}(0.5)$.

\subsubsection{Toeplitz}

The Toeplitz covariance structure assumes each matrix diagonal is equal. Additionally, for our simulations a time $t^\star$ is chosen as the maximum time for which a node is correlated with itself. In practice we choose $t^\star = \frac{7}{10} T$. For our simulations on the Toeplitz covariance matrix we use the same parameters as the AR simulation.

The within-variable covariance matrix ($\Sigma_v$) and between-variable covariance matrix ($\Sigma_{vw}$) are
{\renewcommand*{\arraystretch}{0.75}
\[\Sigma_{v} = \begin{pmatrix}
\Sigma_v^0 & \textbf{0} \\ \textbf{0} & \Sigma_v^1 \end{pmatrix} \hspace{10mm} \Sigma_{vw} = \begin{pmatrix}
\Sigma_{vw}^0 & \textbf{0} \\ \textbf{0} & \Sigma_{vw}^1 \end{pmatrix} \]
}
for 
{\renewcommand*{\arraystretch}{0.75}
    \[\Sigma_{v}^0 = \begin{pmatrix}
    \sigma_{v}^{2} & \rho_{v}\sigma_{v} & \frac{\rho_{v}}{2}\sigma_{v} & \ldots & \frac{\rho_{v}}{t^\star}\sigma_{v} \\
    & \sigma_{v}^{2} & \rho_{v}\sigma_{v} & \ldots & \frac{\rho_{v}}{t^\star-1}\sigma_{v} \\
    & & \sigma_{v}^{2} & \ldots & \frac{\rho_{v}}{t^\star-2}\sigma_{v} \\
    & & & \ddots & \vdots \\ 
    & & & & \sigma_v^2
    \end{pmatrix} \hspace{10mm} \Sigma_v^1 = \begin{pmatrix}
        \sigma_{v}^{2} & 0 & \ldots & 0 \\
        & \sigma_{v}^{2} & {0} & {0}\\
        &  & \ddots & {0}\\
        & & & \sigma_{v}^{2} \\  
    \end{pmatrix}\]
    \\ \[\Sigma_{vw}^0 = \rho_{vw}\sigma_{v}\sigma_{w} \begin{pmatrix}
    1 & \rho_{v}\rho_{w} & \ldots & \frac{\rho_{v}}{t^\star - 1} \frac{\rho_{w}}{t^\star - 1} \\
    & 1 & \ldots &  \frac{\rho_{v}}{t^\star - 2} \frac{\rho_{w}}{t^\star - 1} \\ 
    & & \ddots & \vdots \\ 
    & & & 1
    \end{pmatrix} \hspace{10mm} \Sigma_{vw}^1 = \rho_{vw}\sigma_{v}\sigma_{w} I_{t-t^\star}\textbf{.}\]
}

\subsection{Simulation Frobenius Error}

\begin{figure}[h]
    \centering
    \begin{subfigure}{0.5\textwidth}
        \centering
        \includegraphics[width=2.5in,height=\textheight]{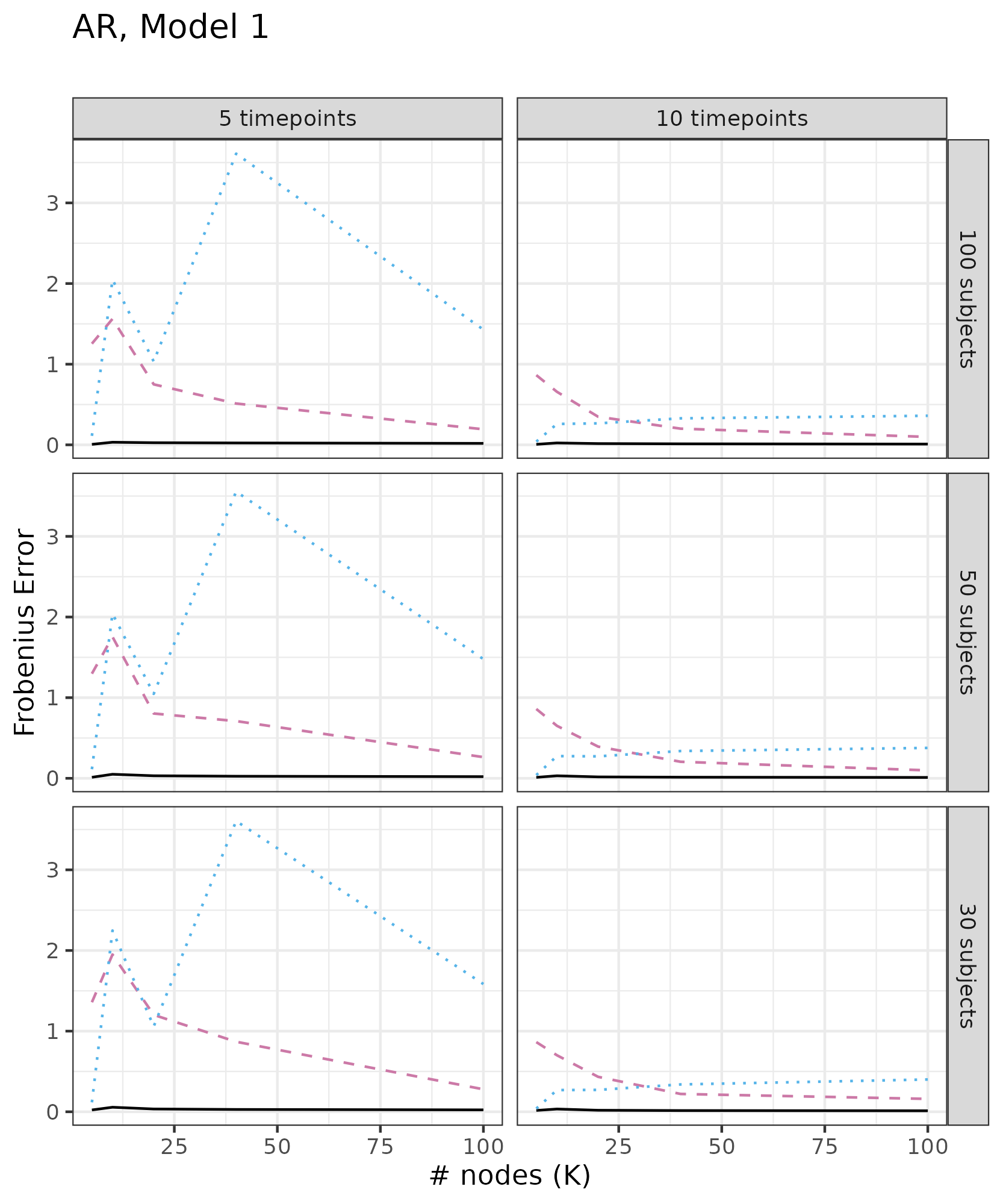}
    \end{subfigure}%
    \begin{subfigure}{0.5\textwidth}
        \centering
        \includegraphics[width=3in,height=\textheight]{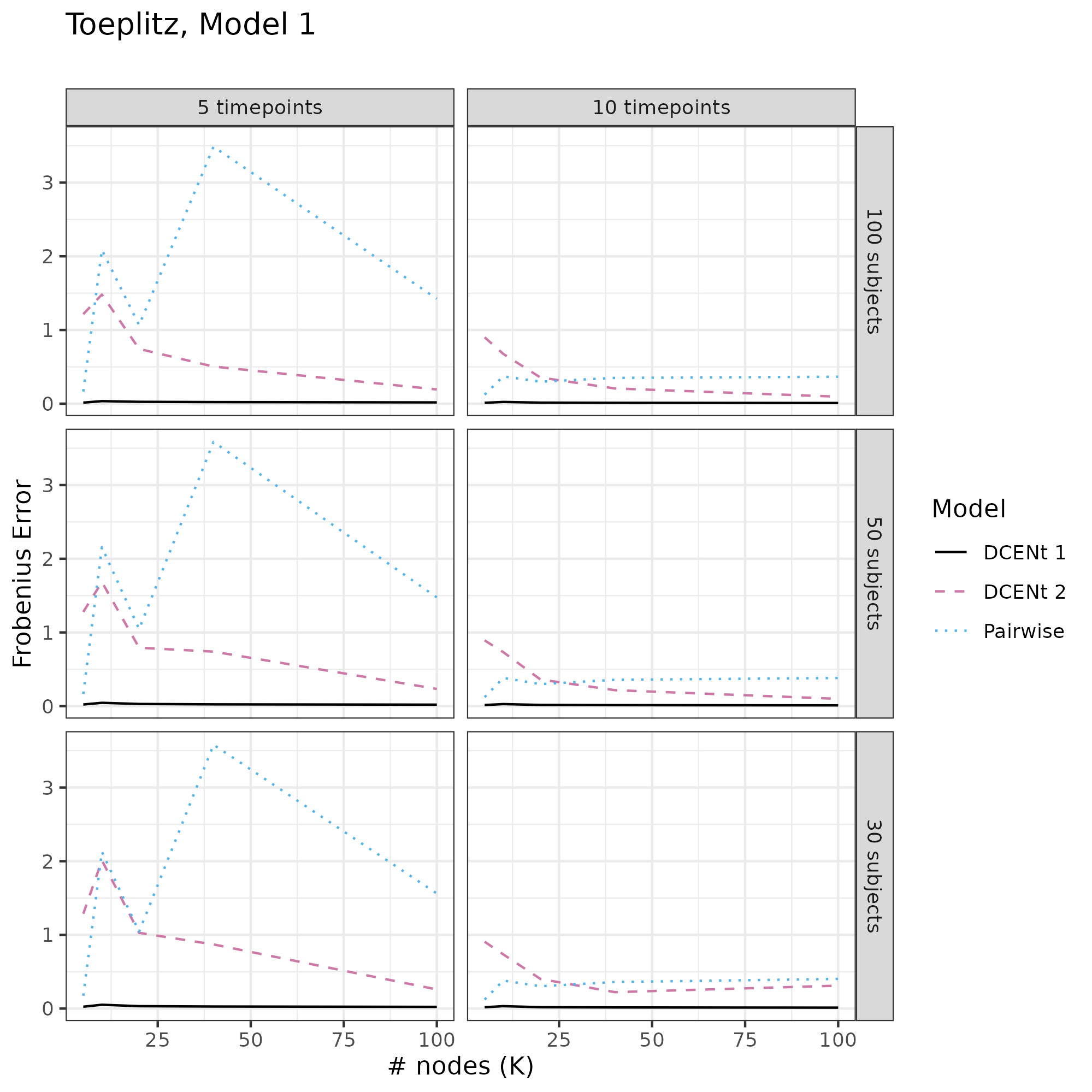}
    \end{subfigure}
    \caption{The Frobenius error in estimating the overall covariance for each of the three algorithms, AR and Toeplitz true covariance matrices with data generation model 1. Panels are split between number of timepoints and number of subjects, increasing from bottom left to top right.}
    \label{fig:V_frob_2}
\end{figure}

\begin{figure}[h]
    \centering
    \begin{subfigure}{0.5\textwidth}
        \centering
        \includegraphics[width=2.5in,height=\textheight]{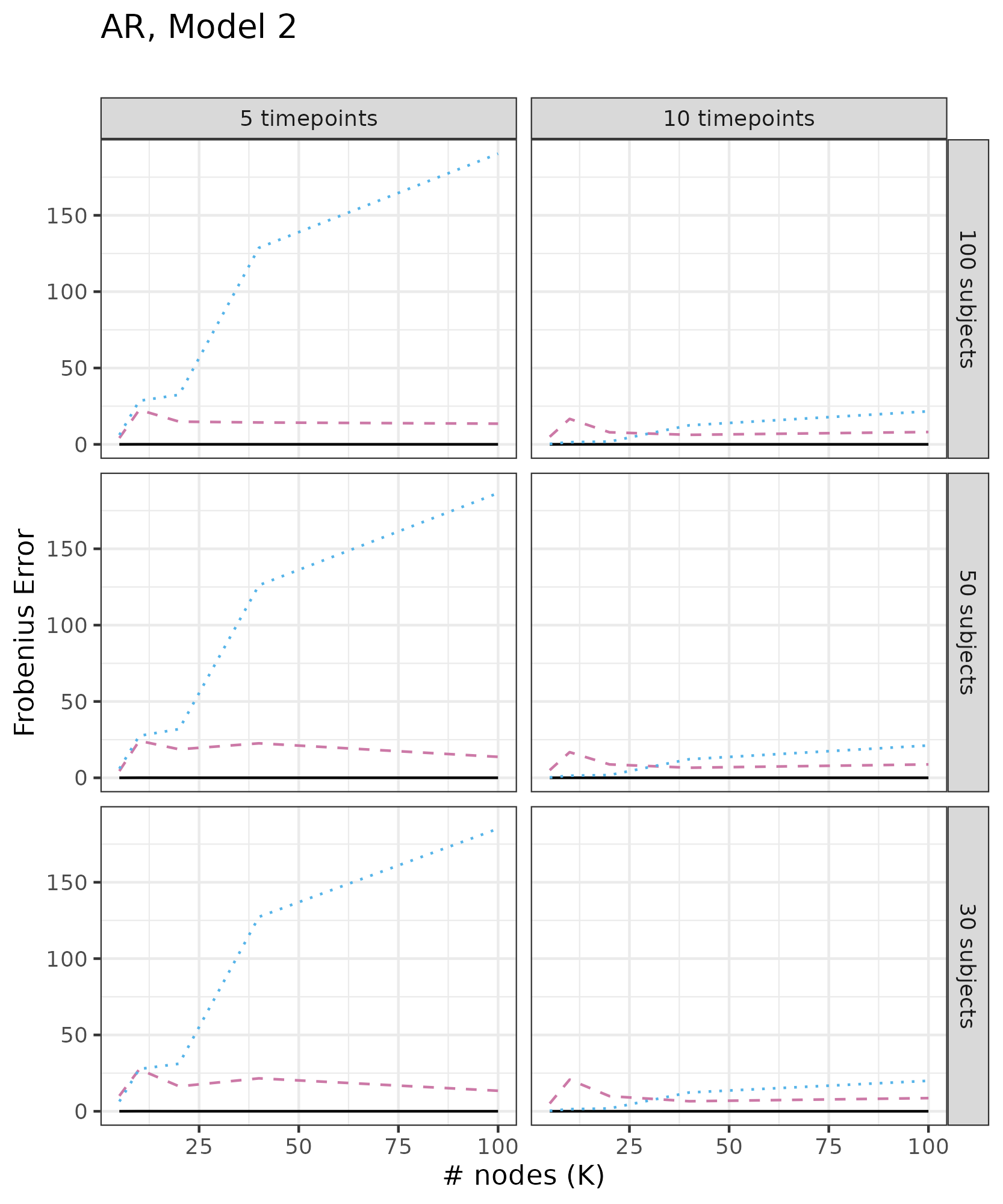}
    \end{subfigure}%
    \begin{subfigure}{0.5\textwidth}
        \centering
        \includegraphics[width=3in,height=\textheight]{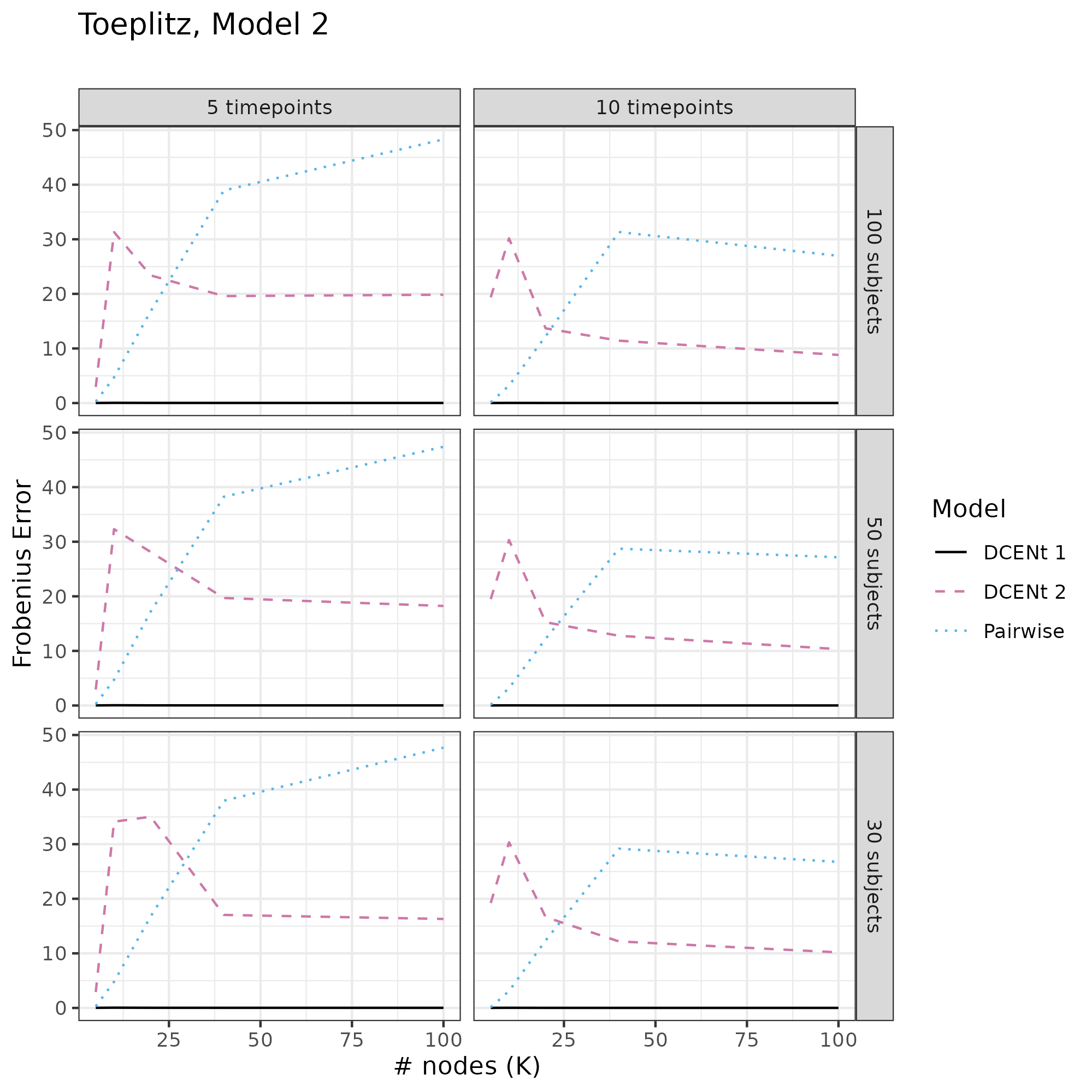}
    \end{subfigure}
    \caption{The Frobenius error in estimating the overall covariance for each of the three algorithms, AR and Toeplitz true covariance matrices with data generation model 2. Panels are split between number of timepoints and number of subjects, increasing from bottom left to top right.}
    \label{fig:V_frob_3}
\end{figure}
\FloatBarrier

\subsection{Simulation $L_1$ Error}
\FloatBarrier

\begin{figure}[h]
    \centering
    \begin{subfigure}{0.5\textwidth}
        \centering
        \includegraphics[width=2.5in,height=\textheight]{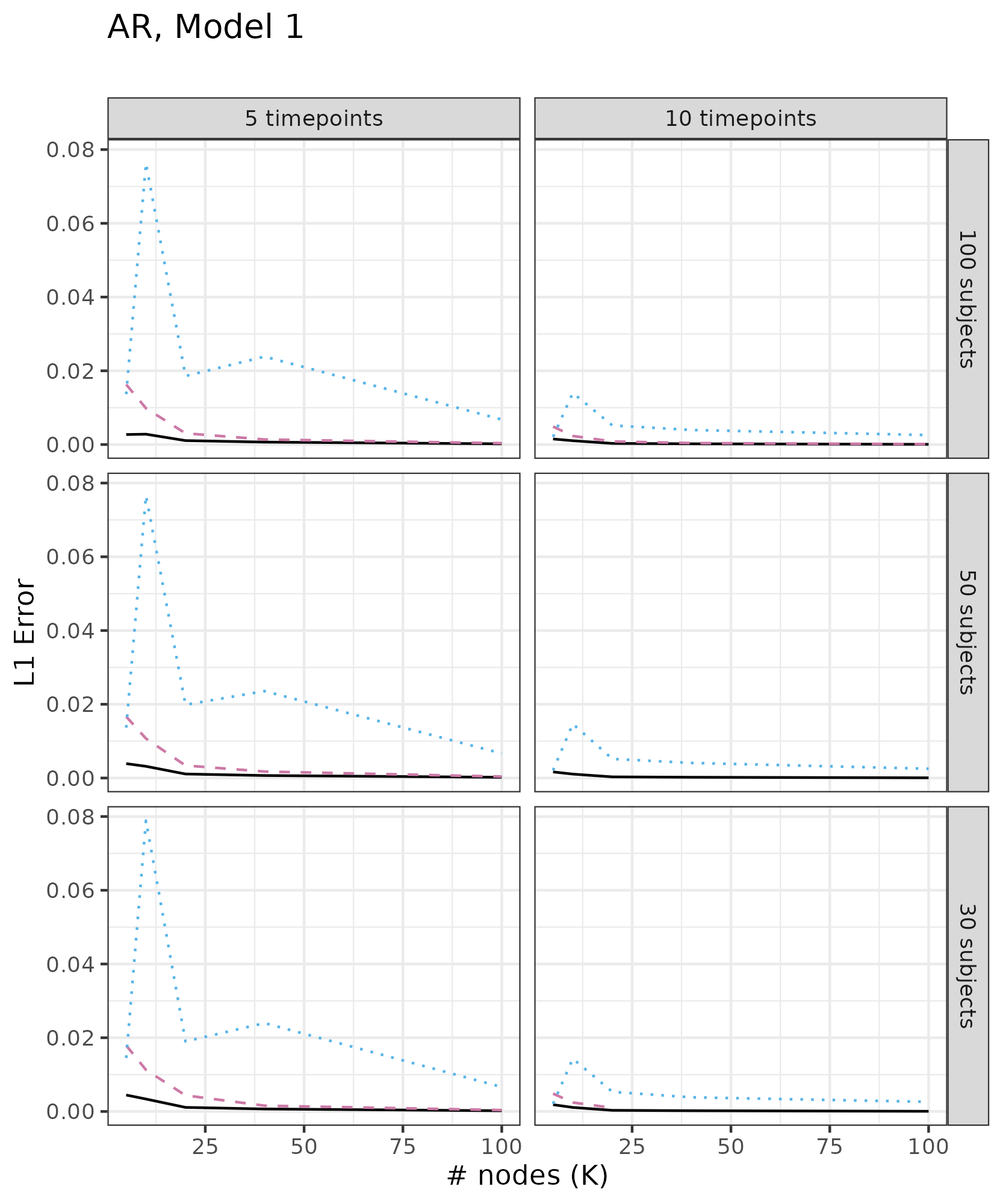}
    \end{subfigure}%
    \begin{subfigure}{0.5\textwidth}
        \centering
        \includegraphics[width=3in,height=\textheight]{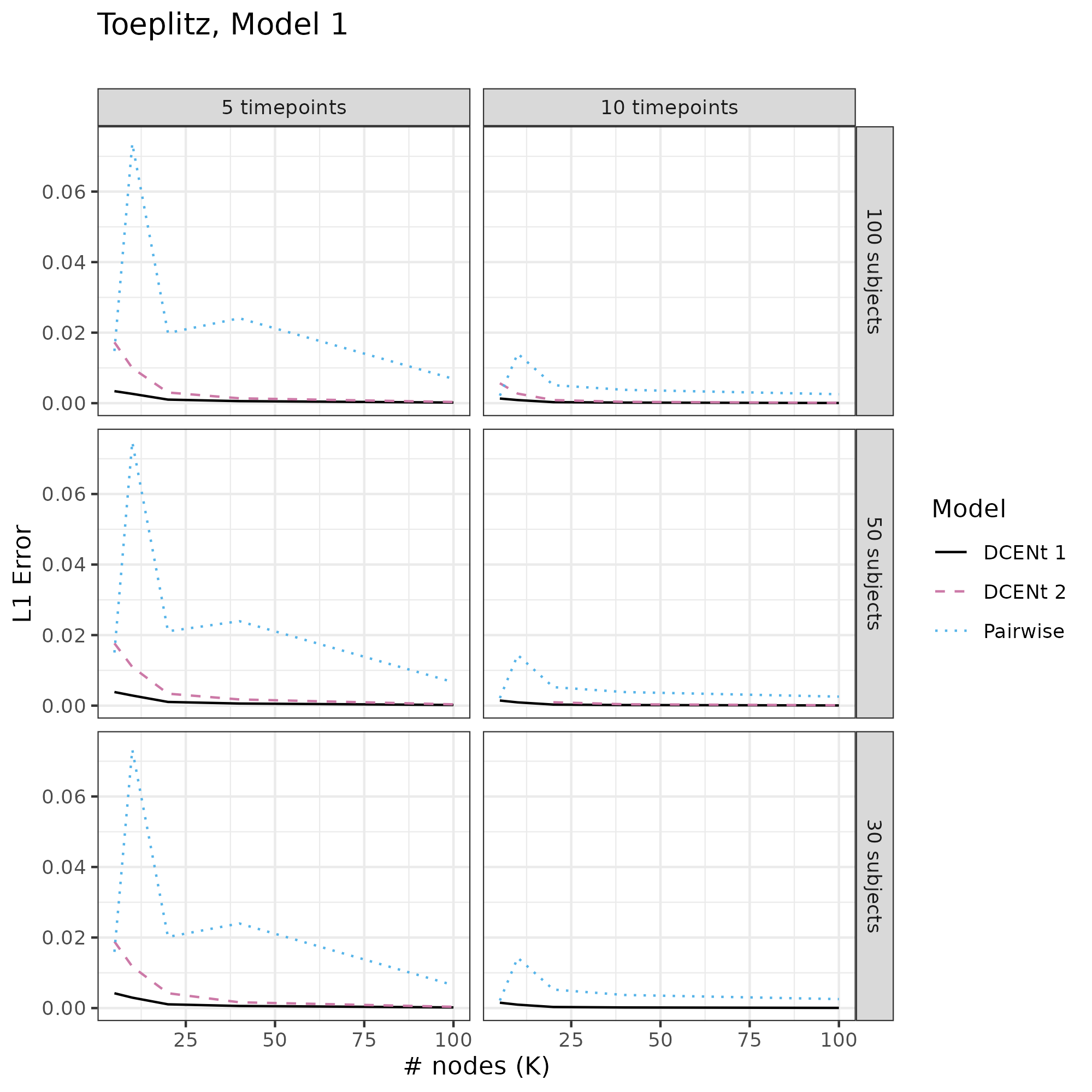}
    \end{subfigure}
    \caption{The $L_1$ error in estimating the overall covariance for each of the three algorithms, AR and Toeplitz true covariance ma''trices with data generation model 1. Panels are split between number of timepoints and number of subjects, increasing from bottom left to top right.}
    \label{fig:V_l1_2}
\end{figure}

\begin{figure}[h]
    \centering
    \begin{subfigure}{0.5\textwidth}
        \centering
        \includegraphics[width=2.5in,height=\textheight]{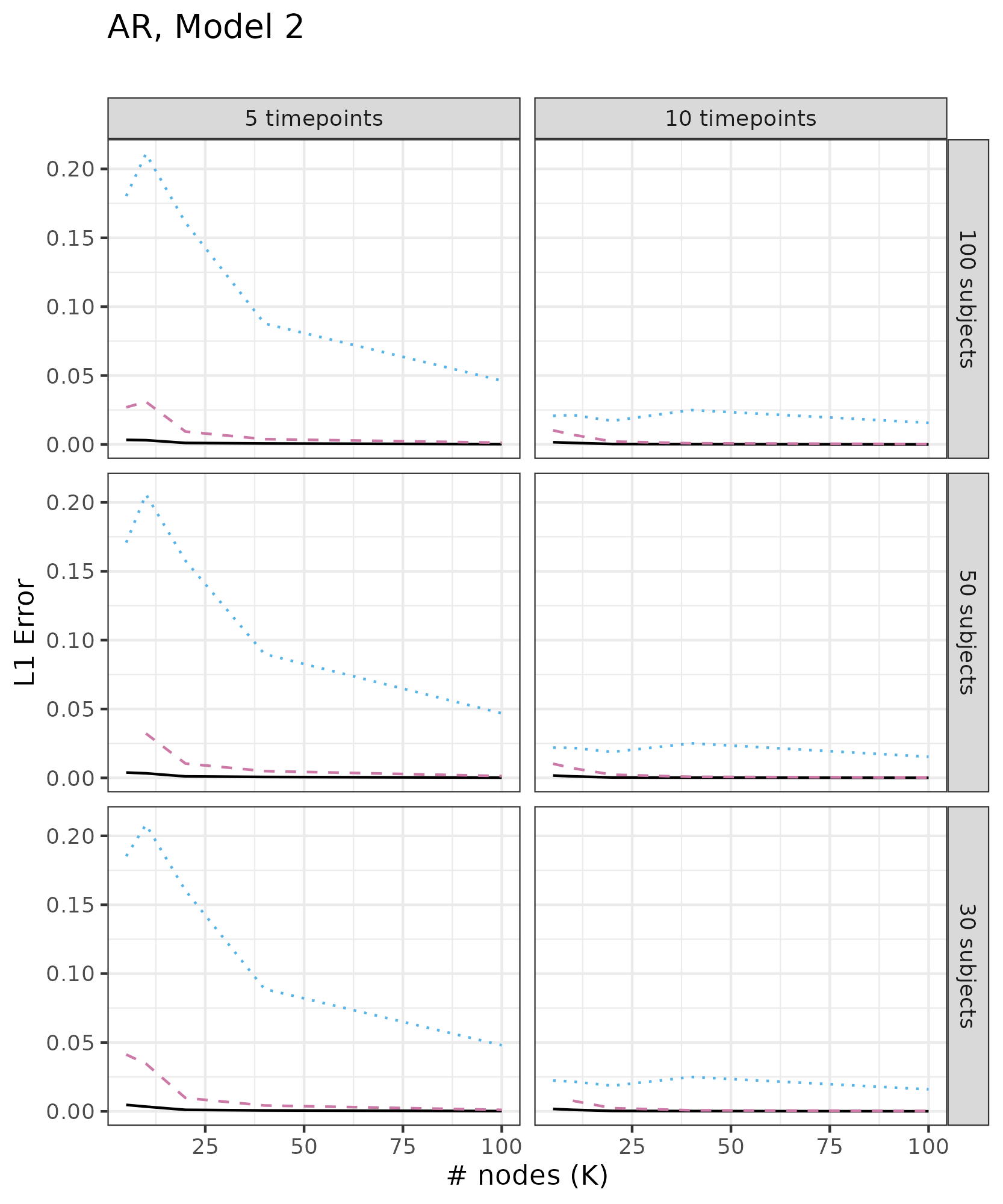}
    \end{subfigure}%
    \begin{subfigure}{0.5\textwidth}
        \centering
        \includegraphics[width=3in,height=\textheight]{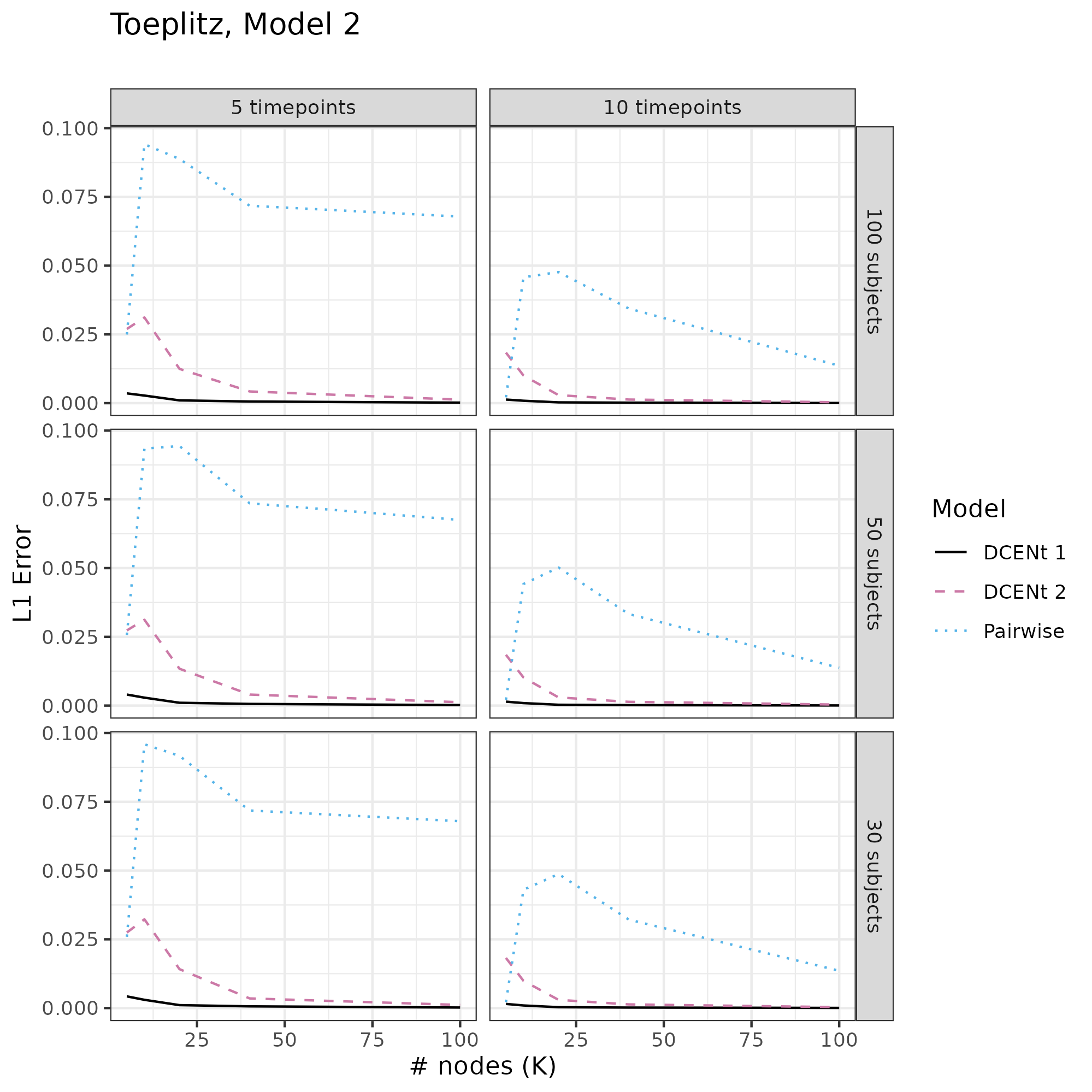}
    \end{subfigure}
    \caption{The $L_1$ error in estimating the overall covariance for each of the three algorithms, AR and Toeplitz true covariance matrices with data generation model 2. Panels are split between number of timepoints and number of subjects, increasing from bottom left to top right.}
    \label{fig:V_l1_3}
\end{figure}

\section{Matrix Multiplication Efficiency Calculations} \label{adx:matrix_calcs}

For even moderately large datasets (50 samples, 20 nodes, 5 timepoints) the matrix multiplication necessary to estimate $\varepsilon$ for DCENt 2 becomes prohibitively slow. Luckily, the block structure of the matrices allows us to break down these operations into manageable chunks. 

We start by splitting the Z matrix into blocks with one block per subject: $Z \coloneqq \begin{bmatrix}Z_1 \\ \vdots \\ Z_n \end{bmatrix}$. Next, we break down the calculation of the matrix $M$:
\begin{align*}
    M &= \Lambda_D^{-1} \left( \begin{bmatrix}
        Z_1^T & ... & Z_n^T
    \end{bmatrix} \begin{bmatrix}
        Z_1 \\ \vdots \\ Z_n
    \end{bmatrix}\right)^{-1} \begin{bmatrix}
        Z_1^T & ... & Z_n^T
    \end{bmatrix}
    \\ &= \Lambda_D^{-1} \left( Z_1^TZ_1 + ... + Z_n^TZ_n \right)^{-1} \begin{bmatrix}
        Z_1^T & ... & Z_n^T
    \end{bmatrix}
    \\ B &\coloneqq \left( Z_1^TZ_1 + ... + Z_n^TZ_n \right)^{-1}
    \\ M &= \begin{bmatrix}
        \Lambda_D^{-1} B Z_1^T & ... & \Lambda_D^{-1} B Z_n^T 
    \end{bmatrix}
\end{align*}
The most significant bottleneck is the calculation of the inverse $(\Lambda_{E_{nt}}^TM^TM\Lambda_{E_{nt}} + I_{nkt})^{-1}$. This is a $nkt \times nkt$ matrix, so inverting it becomes very slow at even relatively small sample sizes. Instead, we use the Woodbury matrix identity to instead invert two $2k \times 2k$ matrices. 
\begin{align*}
    A^{-1} \coloneqq& (\Lambda_{E_{nt}}^TM^TM\Lambda_{E_{nt}} + I_{nkt})^{-1} 
    \\ &= (\Lambda_{E_{nt}}^TZB^T\Lambda_D^{-T}\Lambda_D^{-1}BZ^T\Lambda_{E_{nt}} + I_{nkt})^{-1}
    \\ &= I_{nkt} - \Lambda_{E_{nt}}^TZ((B^T\Lambda_D^{-T}\Lambda_D^{-1}B)^{-1} + Z^T\Lambda_{E_{nt}}\Lambda_{E_{nt}}^TZ)^{-1}Z^T\Lambda_{E_{nt}}
\end{align*}

There is also a bottleneck of multilplying this $nkt \times nkt$ matrix $A^{-1}$ by the $nkt \times nkt$ matrices $\Lambda_{E_{nt}}$ and then multiplying this by the $nkt \times nkt$ matrix $M^TM$. We break these down by taking advantage of the block structure of $Z$. First, we define the following variables to ease the notation: 
\begin{align*}
    r &\coloneqq y - X\beta 
    \\  H &\coloneqq \Lambda_{E_{nt}}Z((B^T\Lambda_D^{-T}\Lambda_D^{-1}B)^{-1} + Z^T\Lambda_{E_{nt}}\Lambda_{E_{nt}}Z)^{-1} = \begin{bmatrix}
        H_1 \\ \vdots \\ H_n
    \end{bmatrix}
\end{align*}
Next, we break down the $\hat\varepsilon$ calculation:
\begin{align*}
    \hat{\varepsilon} &= \Lambda_{E_{nt}} (\Lambda_{E_{nt}}^TM^TM\Lambda_{E_{nt}} + I_{nkt})^{-1}\Lambda_{E_{nt}}^TM^TM r
    \\  &= \Lambda_{E_{nt}} (\Lambda_{E_{nt}} M^TM - \Lambda_{E_{nt}}Z((B^TD^{-T}B)^{-1} + Z^T\Lambda_{E_{nt}}I_{nkt}\Lambda_{E_{nt}}Z)^{-1} Z^T\Lambda_{E_{nt}} \Lambda_{E_{nt}} M^TM) r
    \\ &= \Lambda_{E_{nt}} (\Lambda_{E_{nt}}^T(\Lambda_D^{-1} (Z^TZ)^{-1} Z^T)^T\Lambda_D^{-1} (Z^TZ)^{-1} Z^T\Lambda_{E_{nt}} + I_{nkt})^{-1} \\ &\hspace{5mm}\times \Lambda_{E_{nt}}^T(\Lambda_D^{-1} (Z^TZ)^{-1} Z^T)^T\Lambda_D^{-1} (Z^TZ)^{-1} Z^T r
    \\ &= \Lambda_{E_{nt}} (\Lambda_{E_{nt}}^T Z (Z^TZ)^{-T} \Lambda_D^{-T}\Lambda_D^{-1} (Z^TZ)^{-1} Z^T\Lambda_{E_{nt}} + I_{nkt})^{-1}\\ &\hspace{5mm}\times\Lambda_{E_{nt}}^T(Z (Z^TZ)^{-T}\Lambda_D^{-T})(\Lambda_D^{-1} (Z^TZ)^{-1} Z^T) r
    \\ &= \Lambda_{E_{nt}} (I_{nkt} - \Lambda_{E_{nt}}Z((B^T\Lambda_D^{-T} \Lambda_D^{-1}B)^{-1} + Z^T\Lambda_{E_{nt}}\Lambda_{E_{nt}}Z)^{-1}Z^T\Lambda_{E_{nt}})\\ &\hspace{5mm}\times\Lambda_{E_{nt}}^T(Z (Z^TZ)^{-T}\Lambda_D^{-T})(\Lambda_D^{-1} (Z^TZ)^{-1} Z^T) r
    \\ &= \Lambda_{E_{nt}} (I_{nkt} - H Z^T\Lambda_{E_{nt}})\Lambda_{E_{nt}}^TZ B^T\Lambda_D^{-T}\Lambda_D^{-1} B Z^T r
    \\ &= \Lambda_{E_{nt}} (\Lambda_{E_{nt}}^TZ - H Z^T\Lambda_{E_{nt}}\Lambda_{E_{nt}}^TZ) B^T\Lambda_D^{-T}\Lambda_D^{-1} B Z^T r
    \\ &= \Lambda_{E_{nt}} \left( \begin{bmatrix}
        \Lambda_{E_{t}} Z_1 \\ \vdots \\ \Lambda_{E_{t}} Z_n
    \end{bmatrix} - \begin{bmatrix}
    H_1 \left(\sum_{i=1}^n Z_i^T \Lambda_{E_{t}}\Lambda_{E_{t}} Z_i \right) \\ \vdots  \\ 
    H_n \left(\sum_{i=1}^n Z_i^T \Lambda_{E_{t}}\Lambda_{E_{t}} Z_i \right) 
    \end{bmatrix}  \right) B^T\Lambda_D^{-T}\Lambda_D^{-1} B Z^T r
    \\ &= \Lambda_{E_{nt}} \begin{bmatrix}
        \Lambda_{E_{t}} Z_1  - H_1 \left(\sum_{i=1}^n Z_i^T \Lambda_{E_{t}}\Lambda_{E_{t}} Z_i\right)\\ \vdots \\ \Lambda_{E_{t}} Z_n - H_n \left(\sum_{i=1}^n Z_i^T \Lambda_{E_{t}}\Lambda_{E_{t}} Z_i \right)
    \end{bmatrix}  B^T\Lambda_D^{-T}\Lambda_D^{-1} B \begin{bmatrix}
        Z_1^T & ... & Z_n^T
    \end{bmatrix} r
    \\ &=  \begin{bmatrix}
        \Lambda_{E_{t}}\left(\Lambda_{E_{t}} Z_1  - H_1 \left(\sum_{i=1}^n Z_i^T \Lambda_{E_{t}}\Lambda_{E_{t}} Z_i\right)\right)\\ \vdots \\ \Lambda_{E_{t}}\left(\Lambda_{E_{t}} Z_n - H_n \left(\sum_{i=1}^n Z_i^T \Lambda_{E_{t}}\Lambda_{E_{t}} Z_i\right) \right)
    \end{bmatrix}  B^T\Lambda_D^{-T}\Lambda_D^{-1} B \sum_{i=1}^n Z_i^T r_i
\end{align*}
This transforms the four $nkt \times nkt$ matrix multiplication operations into a series of multiplications with dimensions less than or equal to $kt$.

\end{document}